\documentclass[a4paper]{article}

\usepackage{amsthm}
\usepackage{amsmath}
\usepackage{amssymb}
\usepackage{pgf}
\usepackage{tikz}
\usepackage{xspace}
\usepackage[basic,classfont=caps]{complexity}
\usepackage{todonotes}
\usepackage[inline]{enumitem}
\usepackage{authblk}

\theoremstyle{remark}
\newtheorem{example}{Example}
\newtheorem{remark}{Remark}
\theoremstyle{definition}
\newtheorem{definition}{Definition}
\newtheorem{lemma}{Lemma}
\newtheorem{theorem}{Theorem}
\newtheorem{proposition}{Proposition}
\newtheorem{corollary}{Corollary}

\usepackage{tikz}
\usetikzlibrary{positioning,arrows,automata,shapes} 
\tikzset{->,>=stealth',
every loop/.style={looseness=6},
initial text={},
auto,node distance=1cm,
el/.style={font=\scriptsize},
every state/.style={font=\scriptsize,inner
    sep=0.1mm,minimum size=0.5cm,outer sep=1pt},
psplit/.style={rectangle,fill=black,minimum size=1mm,inner sep=0mm},
smallstate/.style={circle,fill=black,minimum size=1mm,inner sep=0mm},
}

\makeatletter
\let\epsilon\varepsilon
\let\phi\varphi
\let\emptyset\varnothing
\let\rho\varrho
\makeatother

\newcommand{\ie}{\textit{i.e.}\xspace}
\newcommand{\eg}{\textit{e.g.}\xspace}
\newcommand{\calA}{\mathcal{A}}
\newcommand{\calB}{\mathcal{B}}
\newcommand{\calC}{\mathcal{C}}

\newcommand{\calG}{\mathcal{G}}
\newcommand{\calM}{\mathcal{M}}
\newcommand{\dist}[1]{\mathbb{D}(#1)}
\newcommand{\st}{\mathrel{\mid}}
\newcommand{\val}{\mathrm{Val}}
\newcommand{\reach}[2]{\mathbf{Reach}(#2,#1)}
\newcommand{\zero}[1]{\mathbf{Z}_{#1}}
\newcommand{\win}[2]{\mathbf{Win}^{#1}_{#2}}
\newcommand{\prob}{\mathbb{P}}
\newcommand{\pow}{\mathcal{P}}
\newcommand{\reachprob}[3]{\prob^{#2}_{#1}[\Diamond{#3}]}
\newcommand{\until}{\mathrel{\mathsf{U}}}
\newcommand{\untilprob}[4]{\prob^{#2}_{#1}%
    \left[#3 \until #4\right]}
\newcommand{\arenatomdp}[2]{#1_{#2}}
\newcommand{\arenatochain}[3]{#1_{#2}^{#3}}
\newcommand{\eqclass}[1]{\tilde{#1}}
\newcommand{\quo}[1]{{#1}_{\mathord{/}\sim}}

\newcommand{\startmyproof}[1][]{%
\ifthenelse{\equal{#1}{}}{\begin{proof}}{\begin{proof}[Proof #1]}%
}
\newcommand{\stopmyproof}{\end{proof}}

\begin{document}

\title{The Complexity of Graph-Based Reductions for\\Reachability in 
Markov Decision Processes}

\author{St\'ephane Le Roux}
\affil{Department of Mathematics, Technische Universit\"at Darmstadt\\
    \texttt{leroux@mathematik.tu-darmstadt.de}}

\author{Guillermo A. P\'erez}
\affil{Universit\'e libre de Bruxelles, Departement d'Informatique\\
    \texttt{gperezme@ulb.ac.be}}

\maketitle

\begin{abstract}
    We study the never-worse relation (NWR) for Markov decision processes
    with an infinite-horizon reachability objective. A state $q$ is never worse
    than a state $p$ if the maximal probability of reaching the target set of
    states from $p$ is at most the same value from $q$, regardless of the
    probabilities labelling the transitions. Extremal-probability states, end
    components, and essential states are all special cases of the equivalence
    relation induced by the NWR. Using the NWR, states in the same equivalence
    class can be collapsed. Then, actions leading to sub-optimal states can
    be removed. We show that the natural decision problem associated to computing the
    NWR is \coNP-complete. Finally, we extend a previously known incomplete
    polynomial-time iterative algorithm to under-approximate the NWR.
\end{abstract}

\section{Introduction}
Markov decision processes (MDPs) are a useful model for decision-making in the
presence of a stochastic environment. They are used in several fields, including
robotics, automated control, economics, manufacturing and in particular
planning~\cite{rn10}, model-based reinforcement learning~\cite{sll09}, and
formal verification~\cite{bk08}.  We elaborate on the use of MDPs and the need
for graph-based reductions thereof in verification and reinforcement learning
applications below.

Several verification problems for MDPs reduce to
reachability~\cite{bk08,cbgk08}. For instance, MDPs can be model checked against
linear-time objectives (expressed in, say, LTL) by constructing an
omega-automaton recognizing the set of runs that satisfy the objective
and considering the
product of the automaton with the original MDP~\cite{cy95}. In this product MDP,
accepting end components --- a generalization of strongly connected
components --- are identified and selected as target components. The
question of maximizing the probability that the MDP behaviours satisfy the
linear-time objective is thus reduced to maximizing the probability of reaching
the target components.

The maximal reachability probability is computable in polynomial time by
reduction to linear programming~\cite{cy95,bk08}. In practice, however, most model
checkers use value iteration to compute this value~\cite{prism,storm}. The
worst-case time complexity of value iteration is pseudo-polynomial.
Hence, when implementing model checkers it is usual for a graph-based
pre-processing step to remove as many unnecessary states and transitions as
possible while preserving the maximal reachability probability. Well-known
reductions include the identification of extremal-probability states and maximal
end components~\cite{cbgk08,bk08}. The intended outcome of this pre-processing
step is a reduced amount of transition
probability values that need to be considered when
computing the number of iterations required by value iteration.

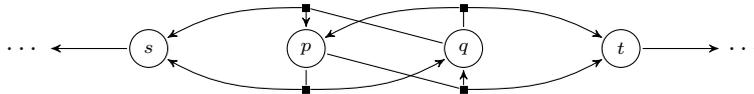
\begin{figure}
	\centering
    \begin{tikzpicture}[node distance=1.5cm]
		\node[state](p){$p$};
		\node[state,right= of p](q){$q$};
		\node[state,left= of p](s){$s$};
		\node[state,right= of q](t){$t$};
        \node[psplit,above=0.2cm of p](qa){ };
        \node[psplit,below=0.2cm of p](pb){ };
        \node[psplit,above=0.2cm of q](qb){ };
        \node[psplit,below=0.2cm of q](pa){ };
        \node[left=1cm of s](ldots){$\dots$};
        \node[right=1cm of t](rdots){$\dots$};

		\path
        (p) edge[-] (pa)
        (p) edge[-] (pb)
        (q) edge[-] (qa)
        (q) edge[-] (qb)
        (pa) edge (q)
        (pa) edge[out=0,in=210] (t)
        (qa) edge (p)
        (qa) edge[out=180,in=30] (s)
        (pb) edge[out=0,in=210] (q)
        (pb) edge[out=180,in=330] (s)
        (qb) edge[out=180,in=30] (p)
        (qb) edge[out=0,in=150] (t)
        (s) edge (ldots)
        (t) edge (rdots)
		;
	\end{tikzpicture}
    \caption{An MDP with states depicted as circles and distributions as
        squares. The maximal reachability probability values from $p$ and $q$
        are the same since, from both,
        one can enforce to reach $s$ with probability $1$, or
        $t$ with probability $1$, using different strategies.}
    \label{fig:selector-gadget}
\end{figure}

The main idea behind MDP reduction heuristics is to identify subsets of states from
which the maximal probability of reaching the target set of states is the same.
Such states are in fact redundant and can be ``collapsed''.
Figure~\ref{fig:selector-gadget} depicts an MDP with actions and
probabilities omitted for clarity. From $p$ and $q$
there are strategies to ensure that $s$ is reached with probability $1$. The
same holds for $t$. For instance, from $p$, to get to $t$ almost surely, one
plays to go to the distribution directly below $q$; from $q$, to the
distribution above $q$. 
Since from the state $p$, there is no
strategy to ensure that $q$ is reached with probability $1$, $p$ and $q$ do not
form an \emph{end component}. In fact, to the best of our knowledge,
no known MDP reduction heuristic captures this example (\ie, recognizes that $p$ and
$q$ have the same maximal reachability probability for all possible values of
the transition probabilities).

In reinforcement learning the actual probabilities labelling the transitions of
an MDP are not assumed to be known in advance. Thus, they have to be estimated
by experimenting with different actions in different states and collecting
statistics about the observed outcomes~\cite{klm96}. In order for the
statistics to be good approximations, the number of experiments has to be high
enough. In particular, when the approximations
are required to be \emph{probably approximately correct}~\cite{valiant13},
the necessary and sufficient number of experiments is
pseudo-polynomial~\cite{ft14}.  Furthermore, the expected number of steps before
reaching a particular state even once may already be exponential (even if all
the probabilities are fixed).  The fact that an excessive amount of experiments
is required is a known drawback of reinforcement
learning~\cite{kawa16,russell2016research}.

A natural and key question to ask in this context is whether the maximal
reachability probability does indeed depend on the actual value of the
probability labelling a particular transition of the MDP. If this is not the
case, then it need not be learnt. One natural way to remove transition
probabilities which do not affect the maximal reachability value is to apply
model checking MDP reduction techniques.

\paragraph*{Contributions and structure of the paper.}
We view the directed graph underlying an MDP as a directed bipartite graph.
Vertices in this graph are controlled by players \emph{Protagonist} and
\emph{Nature}.  Nature is only allowed to choose full-support probability
distributions for each one of her vertices, thus instantiating an MDP from the
graph; Protagonist has strategies just as he would in an MDP. Hence, we consider
infinite families of MDPs with the same support. In the game played between
Protagonist and Nature, and for vertices $u$ and $v$, we are interested in
knowing whether the maximal reachability probability from $u$ is never (in any
of the MDPs with the game as
its underlying directed graph) worse than the same value from $v$.

In Section~\ref{sec:preliminaries} we give the required definitions.  We
formalize the \emph{never-worse relation} in Section~\ref{sec:nwr}. We also show
that we can ``collapse'' sets of equivalent vertices with respect to the NWR
(Theorem~\ref{thm:collapsing}) and remove sub-optimal edges according to the NWR
(Theorem~\ref{thm:opt-trim}). Finally, we also argue that the NWR generalizes
most known heuristics to reduce MDP size before applying linear programming or
value iteration.  Then, in Section~\ref{sec:characterization} we give a
graph-based characterization of the relation
(Theorem~\ref{thm:characterization}), which in turn gives us a \coNP~upper bound
on its complexity. A matching lower bound is presented in
Section~\ref{sec:complexity} (Theorem~\ref{thm:conp-complete}). To conclude, we
recall and extend an iterative algorithm to efficiently (in polynomial time)
under-approximate the never-worse relation from~\cite{ijcai17}. 

\paragraph*{Previous and related work.}
Reductions for MDP model checking were considered in~\cite{djjl01}
and~\cite{cbgk08}. From the reductions studied in both papers,
extremal-probability states, essential states, and end components are computable
using only graph-based algorithms. In~\cite{bccfkkpu14}, learning-based
techniques are proposed to obtain approximations of the
maximal reachability probability in MDPs. Their algorithms, however, do rely on
the actual probability values of the MDP.

This work is also related to the widely studied model of interval MDPs, where
the transition probabilities are given as intervals meant to model the
uncertainty of the numerical values. Numberless MDPs~\cite{fgho14} are a
particular case of the latter in which values are only known to be zero or
non-zero. In the context of numberless MDPs, a special case of 
the question we study can be simply
rephrased as the comparison of the maximal reachability values of two given
states.

In~\cite{ijcai17} a preliminary version of the iterative algorithm we give in
Section~\ref{sec:under-approx} was described, implemented, and shown to be
efficient in practice. Proposition~\ref{pro:bar-rules} was first stated therein.
In contrast with~\cite{ijcai17}, we focus chiefly on characterizing the
never-worse relation and determining its computational complexity. 

\section{Preliminaries}\label{sec:preliminaries}
We use set-theoretic notation to indicate whether a letter $b \in \Sigma$
\emph{occurs} in a word $\alpha = a_0 \dots a_k \in \Sigma^*$, \ie~$b \in \alpha$
if and only if $b = a_i$ for some $0 \le i \le k$.  

Consider a directed graph
$\calG = (V,E)$ and a vertex $u \in V$. We write $uE$ for the set of
\emph{successors} of $u$. That is to say, $uE := \{v \in V \st (u,v) \in E\}$.
We say that a path $\pi = u_0 \dots u_k \in V^*$ in $\calG$ \emph{visits} a
vertex $v$ if $v \in \pi$. We also say that $\pi$ is a $v$--$T$ path,
for $T \subseteq V$, if $u_0 = v$ and $u_k \in T$.

\subsection{Stochastic models}
Let $S$ be a finite set.  We denote by $\dist{S}$ the set of all
\emph{(rational) probabilistic distributions} on $S$, \ie~the set of all
functions $f : S \to \mathbb{Q}_{\ge 0}$ such that $\sum_{s \in S} f(s) = 1$. A
probabilistic distribution $f \in \dist{S}$ has \emph{full support} if $f(s) >
0$ for all $s \in S$.

\begin{definition}[Markov chains]
    A \emph{Markov chain} $\calC$ is a tuple $(Q,\delta)$ where $Q$ is a
    finite set of states and $\delta$ is a
    probabilistic transition function $\delta : Q \to \dist{Q}$.
\end{definition}
A \emph{run} of a Markov chain is a finite non-empty word $\rho = p_0 \dots p_n$
over $Q$. We say $\rho$ \emph{reaches} $q$ if $q = p_i$ for some $0 \le i \le
n$. The \emph{probability of the run} is $\prod_{0 \le i < n}
\delta(p_i,p_{i+1})$.

Let $T \subseteq Q$ be a set of states.  The \emph{probability of (eventually)
reaching $T$} in $\calC$ from $q_0$, which will be 
denoted by $\reachprob{\calC}{q_0}{T}$, is
the measure of the runs of $\calC$ that start at $q_0$ and reach $T$. For
convenience, let us first define the \emph{probability of staying in states from
$S \subseteq Q$ until $T$ is reached}\footnote{$S \until T$
should be read as ``\emph{$S$ until $T$}'' and not
understood as a set union.}, written
$\untilprob{\calC}{q_0}{S}{T}$,
as $1$ if $q_0 \in T$ and otherwise
\[ 
    \sum
    \left\{ \prod_{0 \le i < n} \delta(q_{i},q_{i+1}) \:\middle|\: q_0 \dots q_n
    \in (S \setminus T)^* T \text{ for } n\ge 1 \right\}.
\]
We then define $\reachprob{\calC}{q_0}{T} := \untilprob{\calC}{q_0}{Q}{T}$.

When all runs from $q_0$ to $T$ reach some set $U \subseteq Q$ before, the
probability of reaching $T$ can be decomposed into a finite sum as in the lemma
below.
\begin{lemma}\label{lem:sBt}
    Consider a Markov chain $\calC = (Q,\delta)$, sets of states $U,T \subseteq
    Q$, and a state $q_0 \in Q \setminus U$. If
    $\untilprob{\calC}{q_0}{(Q\setminus U)}{T} = 0$, then
    \[ 
        \reachprob{\calC}{q_0}{T} = \sum_{u \in U}
        \untilprob{\calC}{q_0}{(Q \setminus U)}{u}
        \reachprob{\calC}{u}{T}.
    \]
\end{lemma}

\begin{definition}[Markov decision processes]
    A \emph{(finite, discrete-time) Markov decision process} $\calM$,
    MDP for short, is a tuple $(Q,A,\delta,T)$ where $Q$ is a finite set of
    states, $A$ a finite set of actions, $\delta : Q \times A \to \dist{Q}$ a
    \emph{probabilistic transition function}, and $T \subseteq Q$ a set of
    \emph{target} states.
\end{definition}
For convenience, we write $\delta(q|p,a)$ instead of $\delta(p,a)(q)$.

\begin{definition}[Strategies]
    A \emph{(memoryless deterministic) strategy} $\sigma$ in an MDP $\calM =
    (Q,A,\delta,T)$ is a function $\sigma : Q \to A$.
\end{definition}

Note that we have deliberately defined only memoryless deterministic strategies.
This is at no loss of generality since, in this work, we focus on maximizing the
probability of reaching a set of states. It is known that for this type of
objective, memoryless deterministic strategies suffice~\cite{puterman05}.

\paragraph*{From MDPs to chains.}
An MDP $\calM = (Q,A,\delta,T)$ and
a strategy $\sigma$ induce the Markov chain
$\calM^\sigma = (Q,\mu)$ where $\mu(q) = \delta(q,\sigma(q))$ for all
$q \in Q$.

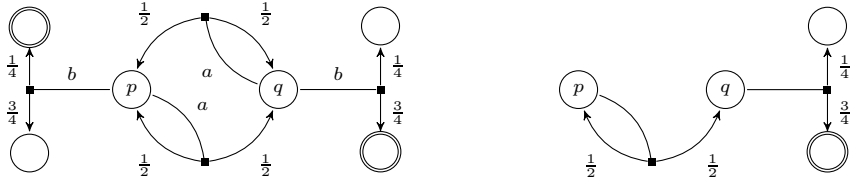
\begin{figure}
	\centering
    \begin{minipage}[b]{0.45\linewidth}
    \centering
	\begin{tikzpicture}
        \node[state,accepting](t1){ };
        \node[psplit,below=0.5cm of t1](pb){ };
        \node[state,below=0.5cm of pb](s1){ };
        \node[state,right= of pb](p){$p$};
        \node[psplit,below right= of p](pa){ };
        \node[state,above right= of pa](q){$q$};
        \node[psplit,above left= of q](qa){ };
        \node[psplit,right= of q](qb){ };
        \node[state,accepting,below=0.5cm of qb](t2){ };
        \node[state,above=0.5cm of qb](s2){ };

        \path
        (pb) edge node[el]{$\frac{1}{4}$} (t1)
        (pb) edge node[el,swap]{$\frac{3}{4}$} (s1)
        (qb) edge node[el,swap]{$\frac{1}{4}$} (s2)
        (qb) edge node[el]{$\frac{3}{4}$} (t2)
        (p) edge[-] node[el,swap]{$b$} (pb)
        (p) edge[bend left,-] node[el]{$a$} (pa)
        (q) edge[-] node[el]{$b$} (qb)
        (q) edge[bend left,-] node[el]{$a$} (qa)
        (pa) edge[bend left] node[el]{$\frac{1}{2}$} (p)
        (qa) edge[bend right] node[el,swap]{$\frac{1}{2}$} (p)
        (qa) edge[bend left] node[el]{$\frac{1}{2}$} (q)
        (pa) edge[bend right] node[el,swap]{$\frac{1}{2}$} (q)
        ;
	\end{tikzpicture}
    \end{minipage}
    \hfill
    \begin{minipage}[b]{0.45\linewidth}
	\centering
	\begin{tikzpicture}
        \coordinate (t1);
        \coordinate[below=0.5cm of t1](pb);
        \coordinate[below=0.5cm of pb](s1);
        \node[state,right= of pb](p){$p$};
        \node[psplit,below right= of p](pa){ };
        \node[state,above right= of pa](q){$q$};
        \coordinate[above left= of q](qa);
        \node[psplit,right= of q](qb){ };
        \node[state,accepting,below=0.5cm of qb](t2){ };
        \node[state,above=0.5cm of qb](s2){ };

        \path
        (qb) edge node[el,swap]{$\frac{1}{4}$} (s2)
        (qb) edge node[el]{$\frac{3}{4}$} (t2)
        (p) edge[bend left,-] node[el]{} (pa)
        (q) edge[-] node[el]{} (qb)
        (pa) edge[bend left] node[el]{$\frac{1}{2}$} (p)
        (pa) edge[bend right] node[el,swap]{$\frac{1}{2}$} (q)
        ;
	\end{tikzpicture}
    \end{minipage}
    \caption{On the left we have an MDP with actions $\{a,b\}$. On the right we
    have the Markov chain induced by the left MDP and the strategy $\{p \mapsto
    a, q \mapsto b\}$.}
    \label{fig:mdp-example}
\end{figure}

\begin{example}
    Figure~\ref{fig:mdp-example} depicts an MDP on the left.
    Circles represent states; double-circles, target states; and
    squares, distributions. The labels on arrows from states to distributions
    are actions; those on arrows from distributions to states, probabilities.

    Consider the strategy $\sigma$ that plays from $p$ the action
    $a$ and from $q$ the action $b$, \ie~$\sigma(p) = a$ and $\sigma(q) = b$.
    The Markov chain on the right is the chain induced by $\sigma$ and the MDP
    on the left. Note that we no longer have action labels.

    The probability of reaching a target state
    from $q$ under $\sigma$ is easily seen to be $3/4$. In other words, if we
    write $\calM$ for the MDP and $T$ for the set of target states then
    $\reachprob{\calM^\sigma}{q}{T} = \frac{3}{4}$.
\end{example}

\subsection{Reachability games against Nature}
We will speak about families of MDPs whose probabilistic transition functions
have the same support. To do so, we abstract away the probabilities and focus on
a game played on a graph. That is, given an MDP $\calM = (Q,A,\delta,T)$ we
consider its \emph{underlying directed graph} $\calG_\calM = (V,E)$ where $V :=
Q \cup (Q\times A)$ and $E := \{ (q,\langle q,a\rangle) \in Q \times (Q \times
A) \} \cup \{ ( \langle p,a \rangle, q) \st \delta(q|p,a) > 0\}$. In
$\calG_\calM$, \emph{Nature} controls the vertices $Q \times A$. We formalize
the game and the \emph{arena} it is played on below.

\begin{definition}[Target arena]
    A \emph{target arena} $\calA$ is a tuple $(V, V_P,E,T)$ such that
    $(V_P,V_N:=V \setminus V_P,E)$ is a bipartite directed graph, $T \subseteq
    V_P$ is a set of \emph{target} vertices, and $uE \neq \emptyset$ for all $u
    \in V_N$.
\end{definition}
Informally, there are two agents in a target arena: \emph{Nature}, who controls
the vertices in $V_N$, and \emph{Protagonist}, who controls
the vertices in $V_P$.

\paragraph*{From arenas to MDPs.}
A target arena $\calA = (V,V_P,E,T)$ together with a family of probability
distributions $\mu = (\mu_u \in \dist{uE})_{u \in V_N}$ induce an MDP. Formally,
let $\arenatomdp{\calA}{\mu}$ be the MDP $(Q,A,\delta,T)$ where $Q = V_P \uplus
\{\bot\}$, $A = V_N$, $\delta(q|p,a)$ is $\mu_a(q)$ if $(p,a),(a,q) \in E$ and
$0$ otherwise, for all $p \in V_P \cup \{\bot\}$
and $a \in A$ we have $\delta(\bot|p,a) = 1$
if $(p,a) \not\in E$.

\paragraph*{The value of a vertex.}
Consider a target arena $\calA = (V,V_P,E,T)$ and a vertex $v \in
V_P$. We define its \emph{(maximal reachability probability) value} with respect
to a family of full-support probability distributions $\mu$ as
\(
    \val^\mu(v) := \max_\sigma \reachprob{\arenatochain{\calA}{\mu}{\sigma}}{v}{T}.
\)
For $u \in V_N$ we set $\val^\mu(u) := \sum\{\mu_u(v) \val^\mu(v) \st v \in
uE\}$.

\section{The never-worse relation}\label{sec:nwr}
We are now in a position to define the relation that we study in this work.
Let us fix a target arena $\calA = (V,V_P,E,T)$.
\begin{definition}[The never-worse relation (NWR)]
    A subset $W \subseteq V$ of vertices is \emph{never worse} than a vertex $v
    \in V$, written $v \unlhd W$, if and only if
    \[ 
        \forall \mu=(\mu_u \in \dist{uE})_{u \in V_N},
        \exists w \in W : \val^\mu(v) \leq \val^\mu(w)
    \]
    where all the $\mu_u$ have \textbf{full support}. We write $v \sim w$ if
    $v \unlhd \{w\}$ and $w \unlhd \{v\}$.
\end{definition}
It should be clear from the definition that $\sim$ is an equivalence relation.
For $u \in V$ let us denote by $\eqclass{u}$ the set of vertices
that are $\sim$-equivalent and belong to the same
owner, \ie~$\eqclass{u}$ is $\{v \in V_P \st v \sim u\}$ if
$u \in V_P$ and $\{v \in V_N \st v \sim u\}$ otherwise.

\begin{figure}
	\centering
    \begin{minipage}[b]{0.45\linewidth}
    \centering
	\begin{tikzpicture}
		\node[state](p){$p$};
		\node[state,below=0.5cm of p](q){$q$};
		\node[state,right= of p](t){$t$};
		\node[state,right= of t,accepting](fin){$\mathit{fin}$};
		\node[state,below=0.5cm of fin](fail){$\mathit{fail}$};

		\path
		(p) edge coordinate[pos=0.5](pa) (t)
		(pa) edge[out=-90,in=45] (q)
        (q) edge[bend right] coordinate[pos=0.5](qa) (t)
		(qa) edge[out=135,in=-45] (p)
		(t) edge coordinate[pos=0.5](ta) (fin)
        (t) edge coordinate[pos=0.5](tb) (fail)
		(ta) edge (fail)
		(tb) edge[bend right] (fin)
		;

		\node[psplit] at (pa) {};
		\node[psplit] at (qa) {};
		\node[psplit] at (ta) {};
		\node[psplit] at (tb) {};
	\end{tikzpicture}
    \end{minipage}
    \hfill
    \begin{minipage}[b]{0.45\linewidth}
	\centering
	\begin{tikzpicture}
		\node[state](p){$p$};
		\node[state,below=0.5cm of p](s){$s$};
		\node[state,right= of p](q){$q$};
		\node[state,right= of s](t){$t$};
		\node[state,right= of q,accepting](fin){$\mathit{fin}$};
		\node[state,right= of t](fail){$\mathit{fail}$};

		\path
        (s) edge coordinate[pos=0.4](sa) (p)
        (s) edge coordinate[pos=0.5](sb) (t)
        (p) edge coordinate[pos=0.5](pa) (q)
		(pa) edge[bend right] (t)
        (q) edge coordinate[pos=0.5](qa) (fin)
		(qa) edge[bend left] (t)
        (t) edge coordinate[pos=0.5](ta) (fail)
		(ta) edge[bend right] (q)
		;

		\node[psplit] at (sa) {};
		\node[psplit] at (sb) {};
		\node[psplit] at (pa) {};
		\node[psplit] at (qa) {};
		\node[psplit] at (ta) {};
	\end{tikzpicture}
    \end{minipage}
    \caption{Two target arenas with $T = \{\mathit{fin}\}$ are shown. Round
    vertices are elements from $V_P$; square vertices, from $V_N$. In the
    left target arena
    we have that $p \unlhd \{q\}$ and $q \unlhd \{p\}$ since any path
    from either vertex visits $t$ before $T$ --- see Lemma~\ref{lem:sBt}. In the
    right target arena we have that $t \unlhd \{p\}$ --- see
    Proposition~\ref{pro:bar-rules}.}
    \label{fig:new-examples}
\end{figure}
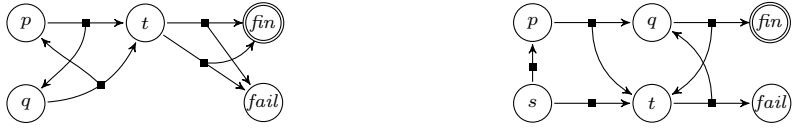

\begin{example}
	Consider the left target arena depicted in Figure~\ref{fig:new-examples}.
    Using Lemma~\ref{lem:sBt}, it is easy to show that neither $p$ nor $q$ is
    ever worse than the other since $t$ is visited before $\mathit{fin}$ by all
    paths starting from $p$ or $q$.
\end{example}

The literature contains various heuristics which consist in computing sets of
states and ``collapsing'' them to reduce the size of the MDP without affecting
the maximal reachability probability of the remaining states. We now show that
we can collapse equivalence classes and, further, remove sub-optimal
distributions using the NWR.

\subsection{The usefulness of the NWR}
We will now formalize the idea of ``collapsing'' equivalent vertices with
respect to the NWR. For convenience, we will also remove self-loops while doing
so.

Consider a target arena $\calA = (V,V_P,E,T)$. We denote by $\quo{\calA}$ its
\emph{$\sim$-quotient}.
That is, $\quo{\calA}$ is the target arena $(S,S_P,R,U)$ where
$S_P = \{ \eqclass{v} \st \exists v \in V_P \}$,
$S = \{ \eqclass{v} \st \exists v \in V_N \} \cup S_P$,
$U = \{ \eqclass{t} \st \exists t \in T\}$, and
\begin{align*}
    R = & \{ (\eqclass{u},\eqclass{v})
        \st
        \exists (u,v) \in \left(V_P \times V_N\right)
        \cap E : vE \setminus \eqclass{u} \neq
        \emptyset\}\\
        \cup & \{(\eqclass{u},\eqclass{v}) \st
        \exists(u,v)\in \left(V_N \times V_P\right) \cap E\}.
\end{align*}
For a family $\mu = (\mu_u \in \dist{uE})_{u \in V_N}$ of full-support
distributions we denote by $\quo{\mu}$ the family $\nu =
(\nu_{\eqclass{u}} \in \dist{\eqclass{u}R})_{\eqclass{u} \in S_N}$
defined as follows. For all $\eqclass{u} \in S_N$
and all $\eqclass{v} \in \eqclass{u}R$ we have
\( 
    \nu_{\eqclass{u}}(\eqclass{v}) = \sum_{w \in \eqclass{v}} \mu_u(w),
\)
where $u$ is any element of $\eqclass{u}$.

The following property of the $\sim$-quotient follows from the fact that
all the vertices in $\eqclass{v}$ have the same maximal probability of
reaching the target vertices.
\begin{theorem}\label{thm:collapsing}
    Consider a target arena $\calA = (V,V_P,E,T)$. For all families $\mu = (\mu_u
    \in \dist{uE})_{u \in V_N}$ of full-support probability distributions and
    all $v \in V_P$ we
    have
    \[
        \max_\sigma \reachprob{\arenatochain{\calA}{\mu}{\sigma}}{v}{T} =
        \max_{\sigma'}
        \reachprob{\arenatochain{\calB}{\nu}{\sigma'}}{\eqclass{v}}{U},
    \]
    where $\calB = \quo{\calA}$, $\nu = \quo{\mu}$, and $U = \{\eqclass{t} \st
    \exists t \in T\}$.
\end{theorem}

We can further remove edges that lead to sub-optimal Nature vertices.
When this is done after $\sim$-quotienting the maximal
reachability probabilities are preserved.
\begin{theorem}\label{thm:opt-trim}
    Consider a target arena $\calA = (V,V_P,E,T)$ such that $\quo{\calA} =
    \calA$. For all families $\mu = (\mu_u \in \dist{uE})_{u \in V_N}$ of
    full-support probability distributions, for all $(w,x) \in E \cap
    \left(V_P \times V_N\right)$
    such that $x \unlhd (wE \setminus \{x\})$, and all $v \in V_P$ we have
    \[ 
        \max_\sigma \reachprob{\arenatochain{\calA}{\mu}{\sigma}}{v}{T} =
        \max_{\sigma'}
        \reachprob{\arenatochain{\calB}{\mu}{\sigma'}}{v}{T},
    \]
    where $\calB = (V,V_P,E \setminus \{(w,x)\},T)$.
\end{theorem}

\subsection{Known efficiently-computable special cases}\label{sec:speccases}
We now recall the definitions of
the set of extremal-probability states, end components, and
essential states. Then, we observe that for all these sets of states 
their maximal probability reachability coincide and their definitions are
independent of the probabilities labelling the transitions of the MDP.
Hence, they are subsets of the set of the equivalence classes induced by $\sim$.

\subsubsection{Extremal-probability states.}
The set of \emph{extremal-probability states} of an MDP $\calM = (Q,A,\delta,T)$
consists of the set of states with maximal probability reachability $0$ and $1$.
Both sets can be computed in polynomial time~\cite{bk08,ch11}. We give below a
game-based definition of both sets inspired by the classical polynomial-time
algorithm to compute them (see, \eg,~\cite{bk08}). Let us fix a target arena
$\calA = (V,V_P,E,T)$ for the sequel.

For a set $T \subseteq V$, let us write $\zero{T} := \{v \in V \st T \text{ is
not reachable from } v \}$.

\paragraph*{(Almost-surely winning) strategies.}
A strategy for Protagonist in a target arena is a function $\sigma : V_P \to
V_N$. We then say that a path $v_0 \dots v_n \in V^*$ is \emph{consistent with
$\sigma$} if $v_i \in V_P \implies \sigma(v_i) = v_{i+1}$ for all
$0 \le i < n$. Let $\reach{\sigma}{v_0}$ denote the set
of vertices reachable from $v_0$ under $\sigma$,
\ie~$\reach{\sigma}{v_0} := \{v_k \st
v_0 \dots v_k \text{ is a path consistent with } \sigma\}$.

We say that a strategy $\sigma$ for Protagonist is \emph{almost-surely
winning from $u_0 \in V$ to $T
\subseteq V_P$} if, after modifying the arena to make all $t \in T$ into sinks,
for all $v_0 \in \reach{\sigma}{u_0}$ we have $\reach{\sigma}{v_0} \cap T \neq
\emptyset$. We denote the set of all such strategies by $\win{v_0}{T}$.

The following properties regarding almost-surely 
winning strategies in a target arena
follow from the correctness of the graph-based algorithm used to compute
extremal-probability states in an MDP~\cite[Lemma 10.108]{bk08}.
\begin{lemma}[From~\cite{bk08}]\label{lem:rel-strat-probs}
    Consider a target arena $\calA = (V,V_P,E,T)$. For all families $\mu =
    (\mu_u \in \dist{uE})_{u \in V_N}$ of full-support probability
    distributions, for all $v \in V_P$ the following hold.
    \begin{enumerate}[label={(\roman*)}]
        \item $\max_\sigma \reachprob{\arenatochain{\calA}{\mu}{\sigma}}{v}{T} = 0
            \iff v \in \zero{T}$
        \item $\forall \sigma : \sigma \in \win{v}{T} \iff
            \reachprob{\arenatochain{\calA}{\mu}{\sigma}}{v}{T} = 1$
    \end{enumerate}
\end{lemma}

\subsubsection{End components.}
Let us consider an MDP $\calM = (Q,A,\delta,T)$. A
set $S \subseteq Q$ of states is an \emph{end component} in $\calM$ if for all
pairs of states $p,q \in S$ there exists a strategy $\sigma$ such that
$\untilprob{\calM^\sigma}{p}{S}{q} = 1$.
\begin{example}
	Let us consider the MDP shown on the left in
    Figure~\ref{fig:mdp-example}.  The set $\{p,q\}$
	is an end component since, by playing $a$ from both states, one can ensure
	to reach either state from the other with probability $1$. 
\end{example}

It follows immediately from the definition of end component that the maximal
probability of reaching $T$ from states in the same end component is the same.
\begin{lemma}\label{lem:probs-in-ec}
    Let $S \subseteq Q$ be an end component in $\calM$. For all $p,q \in S$ we
    have that $\max_\sigma \reachprob{\calM^\sigma}{p}{T} = \max_\sigma
    \reachprob{\calM^\sigma}{q}{T}$.
\end{lemma}

We say an end component is \emph{maximal} if it is maximal with respect to set
inclusion.
Furthermore,
from the definition of end components in MDPs and
Lemma~\ref{lem:rel-strat-probs} it follows 
that we can lift the notion of end component to
target arenas. More precisely, a set $S \subseteq V_P$ is an end component in
$\calA$ if and only if for some family of full-support probability distributions
$\mu$ we have that $S$ is an end component in $\arenatomdp{\calA}{\mu}$ (if and
only if for all $\mu'$ the set $S$ is an end component in
$\arenatomdp{\calA}{\mu'}$).

The set of all maximal end components of a target arena can be computed in
polynomial time using an algorithm based on the strongly connected components of
the graph~\cite{dealfaro97,bk08}.

\subsubsection{Essential states.}
Consider a target arena $\calA = (V,V_P,E,T)$ and
let $\sqsubseteq$ be the smallest relation satisfying the following. For all $u
\in V_P$ we have $u \sqsubseteq u$. For all $u_0,v \in V_P \setminus \zero{T}$
such that $u_0 \neq v$ we have $u_0 \sqsubseteq v$ if for all paths $u_0 u_1
u_2$ we have that $u_2 \sqsubseteq v$ and there is at least one such path.
Intuitively, $u \sqsubseteq v$ holds whenever all paths starting from $u$ reach
$v$. In~\cite{djjl01}, 
the maximal vertices according to $\sqsubseteq$ are called \emph{essential
states}\footnote{This is not the usual notion of essential
states from classical Markov chain theory}.
\begin{lemma}[From~\cite{djjl01}]
    Consider a target arena $\calA = (V,V_P,E,T)$. For all families $\mu =
    (\mu_u \in \dist{uE})_{u \in V_N}$ of full-support probability
    distributions, for all $v \in V_P$ and all essential states $w$,
    if $v \sqsubseteq w$ then
    $\max_\sigma 
    \reachprob{\arenatochain{\calA}{\mu}{\sigma}}{v}{T} = \max_{\sigma'}
    \reachprob{\arenatochain{\calA}{\mu}{\sigma'}}{w}{T}$.
\end{lemma}
Note that, in the left arena in Figure~\ref{fig:new-examples}, $p \sqsubseteq t$
does not hold since there is a cycle between $p$ and $q$ which does not visit
$t$.

It was also shown in~\cite{djjl01} that the
$\sqsubseteq$ relation is computable in polynomial time.

\section{Graph-based characterization of the NWR}\label{sec:characterization}
In this section we give a characterization of the NWR that is reminiscent of the
topological-based value iteration proposed in~\cite{cbgk08}. The main intuition
behind our characterization is as follows. If $v \unlhd W$ does not hold, then
for all $0 < \epsilon < 1$ there is some family $\mu$ of full-support
distributions such that $\val^\mu(v)$ is at least $1-\epsilon$, while
$\val^\mu(w)$ is at most $\epsilon$ for all $w \in W$.  In turn, this must mean
that there is a path from $v$ to $T$ which can be assigned a high probability by
$\mu$ while, from $W$, all paths go with high probability to $\zero{T}$.

We capture the idea of separating a ``good'' $v$--$T$ path from
all paths starting from $W$ by using
partitioning of $V$ into layers $S_i \subseteq V$.
Intuitively, we would like it to be easy to construct a family $\mu$
of probability distributions such that from all vertices in $S_{i+1}$ all paths
going to vertices outside of $S_{i+1}$ end up, with high probability, in lower
layers, \ie~some $S_k$ with $k < i$. A formal definition follows.
\begin{definition}[Drift partition and vertices]
    Consider a 
    target arena $\calA = (V,V_P,E,T)$ and a partition $(S_i)_{0 \le
    i \le k}$ of $V$. For all $0 \le i \le k$, let $S^+_i := \cup_{i < j} S_j$
    and $S^-_i := \cup_{j < i}S_j$, and let $D_i := \{v \in S_i \cap V_N \st vE
    \cap S^-_i \neq \emptyset \}$. We define the set $D:= \cup_{0 <  i < k} D_i$
    of \emph{drift vertices}. The partition is called a \emph{drift partition} if
    the following hold.
    \begin{itemize}
        \item For all $i \le k$ and all $v \in S_i \cap V_P$ we have $vE \cap
            S^+_i = \emptyset$.
        \item For all $i \le k$ and all $v \in S_i \cap V_N$ we have $vE \cap
            S^+_i \neq \emptyset \implies v \in D$.
    \end{itemize}
\end{definition}

Using drift partitions, we can now formalize our characterization of the
negation of the NWR.
\begin{theorem}\label{thm:characterization}
    Consider a 
    target arena $\calA = (V,V_P,E,T)$, a non-empty set of vertices
    $W \subseteq V$, and a vertex $v \in V$. The following are equivalent
    \begin{enumerate}[label={(\roman*)}]
        \item\label{itm:char1} $\lnot \left( v \unlhd W \right)$
        \item\label{itm:char2} There exists a drift partition $(S_i)_{0 \le i
            \le k}$ and a simple path $\pi$ starting in $v$ and ending in $T$ such
            that $\pi \subseteq S_k$ and $W \subseteq S^-_k$.
    \end{enumerate}
\end{theorem}

Before proving Theorem~\ref{thm:characterization} we need an additional
definition and two intermediate results.
\begin{definition}[Value-monotone paths]
    Let $\calA = (V,V_P,E,T)$ be a target arena and consider
    a family of full-support probability
    distributions $\mu = (\mu_u \in \dist{uE})_{u \in V_N}$. A path $v_0
    \dots v_k$ is \emph{$\mu$-non-increasing} if and only if
    $\val^\mu(v_{i+1}) \leq \val^\mu(v_i)$ for all $0 \le i < k$; it is
    \emph{$\mu$-non-decreasing} if and only if $\val^\mu(v_i) \leq
    \val^\mu(v_{i+1})$ for all $0 \le i < k$. 
\end{definition}
It can be shown that from any path in a target arena
ending in $T$ one can obtain a simple non-decreasing one.
\begin{lemma}\label{lem:ndp}
    Consider a target arena $\calA = (V,V_P,E,T)$ and a family of full-support
    probability
    distributions $\mu = (\mu_u \in \dist{uE})_{u \in V_N}$. If there is a path
    from some $v \in V$ to $T$, there is also a simple $\mu$-non-decreasing one.
\end{lemma}

Additionally, we will make use of
the following properties regarding vertex-values.
They formalize the relation between the value of a vertex, its owner, and the
values of its successors.
\begin{lemma}\label{lem:props-val}
    Consider a
    target arena $\calA = (V,V_P,E,T)$ and a family of full-support probability
    distributions $\mu = (\mu_u \in \dist{uE})_{u \in V_N}$.
    \begin{enumerate}[label={(\roman*)}]
        \item\label{itm:prop1} For all $u \in V_P$,
            for all successors $v \in uE$ it holds that
            $\val^\mu(v) \leq \val^\mu(u)$.
        \item\label{itm:prop2} For all $u \in V_N$ it
            holds that
            \[
                (\exists v \in uE : \val^\mu(u) < \val^\mu(v))
                \implies
                (\exists w \in uE : \val^\mu(w) < \val^\mu(u)). 
            \]
    \end{enumerate}
\end{lemma}

\startmyproof[of Theorem~\ref{thm:characterization}]
    Recall that, by definition, $\ref{itm:char1}$ holds if and only if there
    exists a family $\mu = (\mu_u \in \dist{uE})_{u \in V_N}$ of full-support
    probability distributions such that $\forall w \in W: \val^\mu(w) <
    \val^\mu(v)$.

    Let us prove $\ref{itm:char1} \implies \ref{itm:char2}$. Let $x_0 < x_1 <
    \dots$ be the finitely many (\ie at most $|V|$) values that occur in the
    MDP $\arenatomdp{\calA}{\mu}$,
    and let $k$ be such that $ \val^\mu(v) = x_k$. For all $0 \leq i <
    k$ let $S_i := \{u \in V \st \val^\mu(u) = x_i\}$, and let $S_k := V
    \setminus \cup_{i < k}S_i$. Let us show below that the $S_i$ form a drift
    partition.
    \begin{itemize}
        \item $\forall i \leq k, \forall u \in S_i \cap S_P : uE \cap S_i^+=
            \emptyset$ by Lemma~\ref{lem:props-val}.\ref{itm:prop1} (for $i <
            k$) and since $S_k^+ = \emptyset$.
        \item $\forall i \leq k, \forall u \in S_i \cap S_N : uE \cap S_i^+ \neq
            \emptyset \implies x \in D$ by
            Lemma~\ref{lem:props-val}.\ref{itm:prop2} (for $i < k$) and since
            $S_k^+ = \emptyset$.
    \end{itemize}

    We have that $\val^\mu(w) < \val^\mu(v) = x_k$ for all $w \in W$, by
    assumption, so $W \subseteq S_k^-$ by construction. By
    Lemma~\ref{lem:ndp} there exists a simple $\mu$-non-decreasing path $\pi$ from
    $v$ to $T$, so all the vertices occurring in $\pi$ have values at least
    $\val^\mu(v)$, so $\pi \subseteq S_k$.

    We will prove $\ref{itm:char2} \implies \ref{itm:char1}$ by defining some
    full-support 
    distribution family $\mu$. The definition will be partial only, first on
    $\pi \cap V_N$, and then on the drift vertices in $V \setminus S_k$. Let $0
    < \epsilon < 1$, which is meant to be small enough. Let us write $\pi = v_0
    \dots v_{n}$ so that $v_0 = v$ and $v_{n} \in T$.
    Let us define $\mu$ on $\pi \cap V_N$ as follows: for
    all $i < n$, if $v_i \in V_N$ let $\mu_{v_i}(v_{i+1}) := 1-
    \epsilon$. Let $\sigma$ be an arbitrary Protagonist strategy such that for
    all $i < n$, if $v_i \in V_P$ then $\sigma(v_i) := v_{i+1}$.
    Therefore
    \begin{align}
        (1 - \epsilon)^{|V|}  &\leq (1 - \epsilon)^{n} 
        & \text{ since $\pi$ is simple} \nonumber\\
        & \leq \prod_{i < n, v_i \in S_N}\mu_{v_i}(v_{i+1})
        & \text{ by definition of }\mu \nonumber\\
        & \leq \reachprob{\arenatochain{\calA}{\mu}{\sigma}}{v}{T}
        & \nonumber\\
        & \leq \max_{\sigma'}
        \reachprob{\arenatochain{\calA}{\mu}{\sigma'}}{v}{T} = \val^\mu(v). &
        \label{eqn:ineq-v}
    \end{align}
    So, for $0 < \epsilon < 1 - \frac{1}{\sqrt[|V|]{2}}$, we have $\frac{1}{2} <
    (1-\epsilon)^{|V|} \leq \val^\mu(v)$. Below we will further define $\mu$
    such that $\val^\mu(w) \leq  1 - (1-\epsilon)^{|V|} < \frac{1}{2}$ for all
    $w \in W$ and all $0 < \epsilon < 1 - \frac{1}{\sqrt[|V|]{2}}$, which will
    prove $\ref{itm:char2} \implies \ref{itm:char1}$. However, the last part of
    the proof is more difficult.

    For all $1 \leq i \leq k$, for all drift vertices $u \in S_i$, let $\rho(u)$
    be a successor of $u$ in $S_i^-$. Such a $\rho(u)$
    exists by definition of the drift vertices. Then let $\mu_u(\rho(u)) := 1 -
    \epsilon$. We then claim that
    \begin{equation}\label{eqn:ineq-rho}
        \forall u \in D : (1-\epsilon)(1
        -\reachprob{\arenatochain{\calA}{\mu}{\sigma}}{\rho(u)}{T}) \leq 1 -
        \reachprob{\arenatochain{\calA}{\mu}{\sigma}}{u}{T}.
    \end{equation}
    Indeed, $1 - \reachprob{\arenatochain{\calA}{\mu}{\sigma}}{u}{T}$ is the
    probability that, starting at $u$ and following $\sigma$, $T$ is never
    reached; and $(1-\epsilon)(1
    -\reachprob{\arenatochain{\calA}{\mu}{\sigma}}{\rho(u)}{T})$ is the
    probability that, starting at $u$ and following $\sigma$, the second vertex
    is $\rho(u)$ and $T$ is never reached.

    Now let $\sigma$ be an arbitrary strategy, and let us prove the following by
    induction on $j$.
    \[
        \forall 0 \leq j < k, \forall w \in S_j \cup S_j^- :
        \reachprob{\arenatochain{\calA}{\mu}{\sigma}}{w}{T}
        \leq 1 - (1 - \epsilon)^j
    \]

    Base case, $j = 0$: by assumption $W$ is non-empty and included in $S_k^-$,
    so $0 < k$. Also by assumption $T \subseteq S_k$, so $T \cap
    S_0 = \emptyset$. By definition of a drift partition, there are no edges
    going out of $S_0$, regardless of whether the starting vertex is in $V_P$ or
    $V_N$. So there is no path from $w$ to $T$, which implies
    $\val^\mu(w) = 0$ for all $w \in S_0$, and the claim holds for the base
    case. Inductive case, let $w \in S_j$, let $D' := D \cap (S_j \cup S_j^-)$
    and let us argue that every path $\pi$ from $w$ to $T$ must at some point
    leave $S_j \cup S_j^-$ to reach a vertex with higher index, \ie there is
    some edge $(\pi_i,\pi_{i+1})$ from $\pi_i \in S_j \cup S_j^-$ to some
    $\pi_{i+1} \in S_\ell$ with $j < \ell$.  By definition of a drift partition,
    $\pi_i$ must also be a drift vertex, \ie $\pi_i \in D'$. Thus, if we let
    $F:= V_P \setminus D'$, Lemma~\ref{lem:sBt} implies that
    $\reachprob{\arenatochain{\calA}{\mu}{\sigma}}{w}{T}= \sum_{u \in D'}
    \untilprob{\arenatochain{\calA}{\mu}{\sigma}}{w}{F}{u}
    \reachprob{\arenatochain{\calA}{\mu}{\sigma}}{u}{T}$.
    Now, since
    \begin{align*}
        & \sum_{u \in D'}
        \reachprob{\arenatochain{\calA}{\mu}{\sigma}}{u}{T} & \\
        = &
        \sum_{u \in D \cap S^-_j}
        \reachprob{\arenatochain{\calA}{\mu}{\sigma}}{u}{T} + 
        \sum_{u \in D_j}
        \reachprob{\arenatochain{\calA}{\mu}{\sigma}}{u}{T} & \text{by splitting the
        sum}\\
        \leq & \sum_{u \in D \cap S_j^-}
        \reachprob{\arenatochain{\calA}{\mu}{\sigma}}{u}{T} +
        \sum_{u \in D_j}
        (1 - (1-\epsilon)(1 -
        \reachprob{\arenatochain{\calA}{\mu}{\sigma}}{\rho(u)}{T}
        )) & \text{by \eqref{eqn:ineq-rho}}\\
        \leq & \sum_{u \in D \cap S_j^-}
        (1 - (1-\epsilon)^{j-1}) + & \text{by IH and since} \\
        & \sum_{u \in D_j}
        (1 - (1-\epsilon)(1 -\epsilon)^{j-1}) &
        \forall x\in D_j:\rho(x) \in S_j^- \\
        \leq & \sum_{u \in D'}
        (1 - (1-\epsilon)^{j}) & (1-\epsilon)^{j} \leq
        (1-\epsilon)^{j-1}
    \end{align*}
    and $\sum_{u \in D'} \untilprob{\arenatochain{\calA}{\mu}{\sigma}}{w}{F}{u}
    \leq 1$, we have that $\reachprob{\arenatochain{\calA}{\mu}{\sigma}}{w}{T}
    \leq 1 - (1-\epsilon)^j$.
    The induction is thus complete. Since $\sigma$ is arbitrary in the
    calculations above, and since $j < k \leq |V|$, we find that $\val^{\mu}(w)
    \leq 1 - (1-\epsilon)^{|V|}$ for all $w \in W \subseteq S_k^-$.

    For $0 < \epsilon < 1 - \frac{1}{\sqrt[|V|]{2}}$ we have $\frac{1}{2} <
    (1-\epsilon)^{|V|}$, as mentioned after \eqref{eqn:ineq-v}, so $\val^\mu(w)
    \leq 1 - (1-\epsilon)^{|V|} < \frac{1}{2}$.
\stopmyproof

\section{Intractability of the NWR}\label{sec:complexity}
It follows from Theorem~\ref{thm:characterization} that we can decide whether a
vertex is sometimes worse than a set of vertices by guessing a partition of
the vertices and verifying that it is a drift partition. The verification can
clearly be done in polynomial time.
\begin{corollary}
    Given a 
    target arena $\calA = (V,V_P,E,T)$, a non-empty set $W \subseteq V$,
    and a vertex $v \in V$, determining whether $v \unlhd W$ is decidable and in
    \coNP.
\end{corollary}

We will now show that the problem is in fact \coNP-complete already for Markov
chains.
\begin{theorem}\label{thm:conp-complete}
    Given a target arena $\calA = (V,V_P,E,T)$, a non-empty vertex set $W
    \subseteq V$, and a vertex $v \in V$, determining whether $v \unlhd W$ is
    \coNP-complete even if $|uE| = 1$ for all $u \in V_P$.
\end{theorem}
The idea is to reduce the \textsc{$2$-Disjoint Paths problem} (2DP) to the
existence of a drift partition witnessing that $v \unlhd \{w\}$ does not hold,
for some $v \in V$.  Recall that 2DP asks, given a directed graph $\calG =
(V,E)$ and vertex pairs $(s_1,t_1),(s_2,t_2) \in V \times V$, whether there
exists an $s_1$--$t_1$ path $\pi_1$ and an $s_2$--$t_2$ path $\pi_2$ such that
$\pi_1$ and $\pi_2$ are vertex disjoint, \ie~$\pi_1 \cap \pi_2 = \emptyset$.
The problem is known to be \NP-complete~\cite{fhw80,eilam-tzoreff98}. In the
sequel, we assume without loss of generality that
\begin{enumerate*}[label={(\alph*)}]
    \item $t_1$ and $t_2$ are reachable from all $s \in V \setminus
        \{t_1,t_2\}$; and
    \item $t_1$ and $t_2$ are the only sinks $\calG$.
\end{enumerate*}
\startmyproof[of Theorem~\ref{thm:conp-complete}]
    From the 2DP input instance, we construct the target arena $\calA =
    (S,S_P,R,T)$ with $S := V \cup E$, $R := \{ (u,\langle u,v\rangle), (\langle
    u,v\rangle, v) \in S \times S \st (u,v) \in E \text{ or } u = v \in
    \{t_1,t_2\}\}$, $S_P := V \times V$, and $T := \{\langle t_1,t_1 \rangle\}$.
    We will show there are vertex-disjoint $s_1$--$t_1$ and $s_2$--$t_2$ paths
    in $\calG$ if and only if there is a drift partition $(S_i)_{0 \le i \le k}$
    and a simple $s_1$--$t_1$ path $\pi$ such that $\pi \subseteq S_k$ and $s_2
    \in S^-_k$. The result will then follow from
    Theorem~\ref{thm:characterization}.

    Suppose we have a drift partition $(S_i)_{0 \le i \le k}$ with $s_2 \in
    S^-_k$ and a simple path $\pi = v_0 \langle v_0,v_1 \rangle
    \dots \langle v_{n-1},v_n\rangle v_n$ with $v_0 = s_1, v_n = t_1$.
    Since the set $\{t_2,\langle t_2,t_2\rangle\}$ is \emph{trapping} in
    $\calA$, \ie~all paths from vertices in the set visit only vertices from it,
    we can assume that $S_0 = \{t_2,\langle t_2,t_2\rangle\}$. (Indeed,
    for any drift partition, one can obtain a new drift partition by moving any
    trapping set to a new lowest layer.) Now, using
    the assumption that $t_2$ is reachable from all $s \in V \setminus
    \{t_1,t_2\}$ one can show by induction that for all $0 \le j < k$ and for
    all $\rho = u_0 \in S_j$ there is a path $u_0 \dots u_m$ in $\calG$
    with $u_m = t_2$
    and $\rho \subseteq S^-_{j+1}$. This implies that there is a $s_2$--$t_2$
    path $\pi_2$ in $\calG$ such that $\pi_2 \subseteq S^-_k$. It follows that
    $\pi_2$ is
    vertex disjoint with the $s_1$--$t_1$ path $v_0 \dots v_n$ in $\calG$.

    Now, let us suppose that we have $s_1$--$t_1$ and $s_2$--$t_2$ vertex
    disjoint paths $\pi_1 = u_0 \dots u_n$
    and $\pi_2 = v_0 \dots v_m$. Clearly, we can assume
    both $\pi_1,\pi_2$ are simple.
    We will construct a
    partition $(S_i)_{0 \le i \le m + 1}$ and show that it is indeed a drift
    partition,
    that $u_0 \langle u_0,u_1 \rangle \dots \langle u_{n-1},u_n \rangle u_n
    \subseteq S_{m+1}$, and $s_2 = v_0 \in S^-_{m+1}$. Let us
    set $S_0 := \{\langle v_{m-1},v_m \rangle, v_m, \langle t_2,t_2\rangle\}$,
    $S_i := \{\langle v_{m-i-1},v_{m-i}\rangle, v_{m-i}\}$
    for all $0 < i \le m$, and $S_{m+1} := S \setminus \cup_{0 \le i \le m}
    S_i$. Since $\pi_2$ is simple, $(S_i)_{0 \le i \le m+1}$ is a partition of
    $V$. Furthermore, we have that $s_2 = v_0 \in S^-_{m+1}$, and
    $u_0 \langle u_0,u_1 \rangle \dots \langle u_{n-1},u_n \rangle u_n
    \subseteq
    S_{m+1}$ since $\pi_1$ and $\pi_2$ are vertex disjoint.  Thus, it only
    remains for us to argue that for all $0 \le i \le m+1$: for all $w \in S_i
    \cap S_N$ we have $wR \cap S^+_i = \emptyset$, and for all $w \in S_i \cap
    V_N$ we have $wR \cap S^+_i \neq \emptyset \implies wR \cap S^-_i \neq
    \emptyset$.  By construction of the $S_i$, we have that $eR \subseteq S_i$
    for all $0 \le i \le m$ and all $e \in S_i \cap S_P$. Furthermore, for all
    $0 < i \le m$, for all $x \in S_i \cap S_N = \{v_{m-i}\}$, there exists $y
    \in S_{i-1} \cap S_P = \{\langle v_{m-i},v_{m-i+1}\rangle \}$ such that
    $(x,y) \in R$ --- induced by $(v_{m-i},v_{m-1+1}) \in E$ from $\pi_2$. To
    conclude, we observe that since $S_{0} = \{\langle v_{m-1}, v_m\rangle, v_m =
    t_2, \langle t_2,t_2 \rangle\}$ and $\{t_2,\langle t_2,t_2 \rangle\}$
    is trapping in
    $\calA$, the set $t_2R$ is contained in $S_0$.
\stopmyproof

\section{Efficiently under-approximating the NWR}\label{sec:under-approx}
Although the full NWR cannot be efficiently computed for a given MDP, we can
hope for ``under-approximations'' that are accurate and efficiently computable.

\begin{definition}[Under-approximation of the NWR]
    Let $\calA = (V,V_P,E,T)$ be a target arena and consider a relation
    $\mathord{\preceq} : V
    \times \pow(V)$. The relation $\preceq$ is an \emph{under-approximation} of
    the NWR if and only if $\preceq \subseteq \unlhd$.
\end{definition}
We denote by $\preceq^*$ the \emph{pseudo transitive closure} of $\preceq$. That
is, $\preceq^*$ is the smallest relation such that
$\preceq \subseteq \preceq^*$ and 
for all $u \in V, X \subseteq V$ if there exists $W \subseteq V$ such that $u
\preceq^* W$ and $w \preceq^* X$ for all $w \in W$, then $u \preceq^* X$.
\begin{remark}\label{rem:props-trans-closure}
    The empty set is an under-approximation of the NWR.  For all
    under-approximations $\preceq$ of the NWR, the pseudo transitive closure
    $\preceq^*$ of $\preceq$ is also an under-approximation of the NWR.
\end{remark}

In~\cite{ijcai17}, efficiently-decidable sufficient conditions for
the NWR were given. In particular, those conditions suffice to infer relations
such as those in the right MDP from Figure~\ref{fig:new-examples}. We recall
(Proposition~\ref{pro:bar-rules}) and
extend (Proposition~\ref{pro:equiv-rules}) these conditions below.

\begin{proposition}[From~\cite{ijcai17}]\label{pro:bar-rules}
    Consider a target arena $\calA = (V,V_P,E,T)$ and an under-approximation
    $\preceq$ of the NWR. For all vertices $v_0 \in V$, and sets $W \subseteq V$
    the following hold.
    \begin{enumerate}[label={(\roman*)}]
        \item\label{itm:happy-bar}
            If there exists $S \subseteq \{ s \in V \st s \preceq W\}$ such
            that there exists no path $v_0 \dots v_n \in (V \setminus S)^*T$,
            then $v_0 \unlhd W$.
        \item\label{itm:sad-bar}
            If $W = \{w\}$ and there exists $S
            \subseteq \{ s \in V_P \st w \preceq \{s\}\}$ such that $\win{v_0}{S
            \cup T} \neq \emptyset$, then $w \unlhd \{v_0\}$.
    \end{enumerate}
\end{proposition}
\startmyproof[Sketch]
    The main idea of the proof of item~\ref{itm:happy-bar}
    is to note that $S$ is visited
    before $T$. The desired result then follows from Lemma~\ref{lem:sBt}.
    For item~\ref{itm:sad-bar}, we intuitively have that there is a strategy to
    visit $T$ with some probability or visit $W$, where the chances of visiting
    $T$ are worse than before. We then show that it is never worse to start from
    $v_0$ to have better odds of visiting $T$.
\stopmyproof

The above ``rules'' give an iterative algorithm to obtain increasingly better
under-approximations of the NWR: from $\preceq_i$ apply the rules and obtain a
new under-approximation $\preceq_{i+1}$ by adding the new pairs and taking the
pseudo transitive closure; then repeat until convergence.
Using the special cases from Section~\ref{sec:speccases} we can obtain
a nontrivial initial under-approximation $\preceq_0$ of the
NWR in polynomial time.

The main problem is how to avoid testing all
subsets $W \subseteq V$ in every iteration. One natural way to
ensure we do not consider all subsets of vertices in every iteration is to apply
the rules from Proposition~\ref{pro:bar-rules} only on the successors of
Protagonist vertices.

In the same spirit of the iterative algorithm described above, we now give two
new rules to infer NWR pairs.
\begin{proposition}\label{pro:equiv-rules}
    Consider a target arena $\calA = (V,V_P,E,T)$ and $\preceq$
    an under-approximation
    of the NWR.
    \begin{enumerate}[label={(\roman*)}]
        \item\label{itm:equiv-nature}
            For all $u \in V_N$, if for all $v,w \in uE$ we have $v \preceq \{w\}$
            and $w \preceq \{v\}$, then $u \sim x$ for
            all $x \in uE$.
        \item\label{itm:equiv-prot}
            For all $u,v \in V_P \setminus T$, if for all $w \in uE$ such that
            $w \preceq  (uE\setminus\{w\})$ does not hold we have that $w
            \preceq vE$, then $u \unlhd \{v\}$.
    \end{enumerate}
\end{proposition}
\startmyproof[Sketch]
    Item~\ref{itm:equiv-nature} follows immediately from the definition of
    $\val$. For item~\ref{itm:equiv-prot} one can use the Bellman optimality
    equations for infinite-horizon reachability in MDPs to show that since
    the successors of $v$ are never worse than the non-dominated successors of
    $u$, we must have $u \unlhd \{v\}$.
\stopmyproof

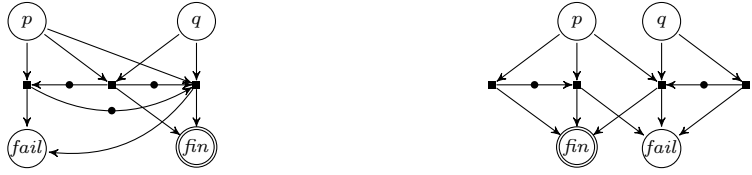
\begin{figure}
	\centering
    \begin{minipage}[b]{0.45\linewidth}
    \centering
	\begin{tikzpicture}
        \node[psplit](u){ };
        \node[psplit,right= of u](v){ };
        \node[psplit,right= of v](z){ };
        \node[state,above=0.5cm of u](p){$p$};
        \node[state,above=0.5cm of z](q){$q$};
		\node[state,below=0.5cm of z,accepting](fin){$\mathit{fin}$};
		\node[state,below=0.5cm of u](fail){$\mathit{fail}$};

		\path
        (p) edge (u)
        (p) edge (v)
        (p) edge (z)
        (q) edge (v)
        (q) edge (z)
        (u) edge (fail)
        (u) edge[bend right] coordinate[pos=0.5](uz) (z)
        (v) edge (fin)
        (v) edge coordinate[pos=0.5](vu) (u)
        (v) edge coordinate[pos=0.5](vz) (z)
        (z) edge (fin)
        (z) edge[bend left] (fail)
        ;

        \node[smallstate] at (uz){};
        \node[smallstate] at (vu){};
        \node[smallstate] at (vz){};
	\end{tikzpicture}
    \end{minipage}
    \hfill
    \begin{minipage}[b]{0.45\linewidth}
	\centering
	\begin{tikzpicture}
        \node[psplit](u){ };
        \node[psplit,right= of u](v){ };
        \node[psplit,right= of v](x){ };
        \node[psplit,right= of x](z){ };
        \node[state,above=0.5cm of v](p){$p$};
        \node[state,above=0.5cm of x](q){$q$};
		\node[state,below=0.5cm of v,accepting](fin){$\mathit{fin}$};
		\node[state,below=0.5cm of x](fail){$\mathit{fail}$};

		\path
        (p) edge (u)
        (p) edge (v)
        (p) edge (x)
        (q) edge (x)
        (q) edge (z)
        (u) edge coordinate[pos=0.5](uv) (v)
        (u) edge (fin)
        (v) edge (fin)
        (v) edge (fail)
        (x) edge (fin)
        (x) edge (fail)
        (z) edge coordinate[pos=0.5](zx) (x)
        (z) edge (fail)
        ;
        \node[smallstate] at (uv) {};
        \node[smallstate] at (zx) {};
	\end{tikzpicture}
    \end{minipage}
    \caption{Two target arenas with $T = \{\mathit{fin}\}$ are shown. Using
    Propositions~\ref{pro:bar-rules} and~\ref{pro:equiv-rules} one can conclude
    that $p \sim q$ in both target arenas.}
    \label{fig:equiv-example}
\end{figure}

The rules stated in Proposition~\ref{pro:equiv-rules}
can be used to infer relations like those depicted
in Figure~\ref{fig:equiv-example} and are clearly seen to be computable in
polynomial time as they speak only of successors of vertices.

\section{Conclusions}
We have shown that the never-worse relation is, unfortunately, not computable in
polynomial time. On the bright side, we have extended the iterative
polynomial-time algorithm from~\cite{ijcai17} to under-approximate the relation.
In that paper, a prototype implementation of the algorithm was used to
empirically show that interesting MDPs (from the set of benchmarks included in
PRISM~\cite{prism}) can be drastically reduced. 

As future work, we believe it would be interesting to implement an exact
algorithm to compute the NWR using SMT solvers. Symbolic implementations of the
iterative algorithms should also be tested in practice. In a more theoretical
direction, we observe that the planning community has also studied maximizing
the probability of reaching a target set of states under the name of
MAXPROB (see, \eg,~\cite{kmwg11,shb16}).
There, online approximations of the NWR would make more sense than the
under-approximation we have proposed here. Finally, one could define a notion of
never-worse for finite-horizon or quantitative
objectives.

\section*{Acknowledgements}
The research leading to these results was supported by the ERC Starting grant
279499: inVEST. Guillermo A. P\'erez is an F.R.S.-FNRS Aspirant and FWA
postdoc fellow.

We thank Nathana\"el Fijalkow for pointing out the relation between this work
and the study of interval MDPs and numberless MDPs.  We also thank Shaull
Almagor, Micha\"el Cadilhac, Filip Mazowiecki, and Jean-Fran\c{c}ois Raskin for
useful comments on earlier drafts of this paper.

\bibliographystyle{abbrv}
\bibliography{refs}

\newpage
\appendix

\section{Preliminaries}
\subsection{Definitions}
\begin{definition}[Memoryful strategies]
    A \emph{(deterministic) strategy} $\sigma$ in an MDP $\calM =
    (Q,A,\delta,T)$ is a function $\sigma : (QA)^*Q \to A$.
\end{definition}
A \emph{finite-memory} strategy $\sigma$ is a strategy that can be encoded as a
deterministic \emph{Mealy machine} $\calA = (S,s_I,Q,A,\lambda_u,\lambda_o)$ where
$S$ is a finite set of (memory) states, $s_I$ is the initial state, $\lambda_u :
S \times Q \to S$ is the update function and $\lambda_o : S \times Q \to A$
is the output function. The machine encodes $\sigma$ in the
following sense: $\sigma(q_0
a_0 \dots q_n) = \lambda_o(s_{n},q_{n})$ where $s_0 = s_I$ and $s_{i+1} =
\lambda_u(s_i,q_i)$ for all $0 \le i < n$.  We then say that $\calA$
\emph{realizes} the strategy $\sigma$ and that $\sigma$ has \emph{memory} $|S|$.
In particular, strategies which have memory $1$ are said to be \emph{positional}
(or \emph{memoryless}).

\subsection{Bellman equations}
The following result~\cite[Theorem 10.100]{bk08} about MDPs will be useful.
\begin{lemma}\label{lem:bellman-ineqs}
    Consider an MDP $\calM = (Q,A,\delta,T)$. The vector $(x_p)_{p \in Q}$ with
    $x_p = \max_\sigma \reachprob{\calM^\sigma}{p}{T}$ is the unique solution of
    the following equation system.
    \begin{itemize}
        \item If $p \in T$ then $x_p = 1$,
        \item if $p \in \zero{T}$ then $x_p = 0$,
        \item otherwise
            \[
                x_p = \max_{a \in A} \sum_{q \in Q} \delta(q|p,a) \cdot x_q.
            \]
    \end{itemize}
\end{lemma}

\section{Proof of Lemma~\ref{lem:sBt}}
\startmyproof
    Since $\untilprob{\calC}{q_0}{(Q\setminus U)}{T} = 0$, then
    $q_0 \not\in T$.
    Thus, by definition, we have
    \begin{equation}\label{eqn:without-one}
        \reachprob{\calC}{q_0}{T} = \sum_{n \ge 1} \left\{ \prod_{0 \le i < n}
        \delta(q_{i},q_{i+1}) \:\middle|\: q_0 \dots q_n \in (Q \setminus T)^* T
        \right\}.
    \end{equation}
    The set of runs that start at $q_0$ and reach $T$ is equivalent to the union
    of: the set of runs that start at $q_0$ and stay in a set $S \subseteq Q$
    until they reach $T$; and the set of runs that start at $q_0$ and
    reach some state from $Q \setminus S$
    before eventually reaching $T$. The measure of the runs from the first set is
    $\untilprob{\calC}{q_0}{S}{T}$. Let us denote by $\prob[\tau_{(Q
    \setminus S)} < \tau_T]$ the measure of the runs from the second set. Since
    in Markov chains we have that for all runs $p_0 \dots p_i \dots p_m$ the
    probabilities of the prefix $p_0 \dots p_i$ and the suffix $p_i \dots p_m$
    are independent, we can rewrite~\eqref{eqn:without-one} as follows.
    \[
        \reachprob{\calC}{q_0}{T} = \untilprob{\calC}{q_0}{(Q \setminus U)}{T} +
        \prob[\tau_{U} < \tau_T]
    \]
    By assumption we have that the first summand is equal to $0$, so
    \[
        \reachprob{\calC}{q_0}{T} = 
        \prob[\tau_{U} < \tau_T].
    \]
    Also by assumption, we know that there are no runs of $\calC$ starting at
    $q_0$ and staying in $U \subseteq Q$ until reaching $T$.  Hence, the set of
    runs from $q_0$ that reach some state in $U$ and then eventually reach $T$
    is exactly the union, over all $u \in U$, of the sets of runs $q_0 \dots
    q_\ell \dots q_n \in Q^*$ that satisfy
    \begin{itemize}
        \item $\forall 0 \le i < \ell: q_i \not\in U$,
        \item $q_\ell = u$,
        \item $\forall \ell < j < n: q_j \not\in T$, and
        \item $q_n \in T$.
    \end{itemize}
    Once more using the fact that events (\ie~transition probabilities) in a
    Markov chain are independent of the history (\ie~run prefixes) we can write
    the measure of the above set as follows.
    \[
        \prob[\tau_U < \tau_T] = \sum_{u \in U}
        \untilprob{\calC}{q_0}{(Q\setminus U)}{u}\reachprob{\calC}{u}{T}
    \]
    This concludes the proof.
\stopmyproof

\section{Proof of Theorem~\ref{thm:collapsing}}
We will first prove that collapsing only the Protagonist vertices is safe.

Consider a target arena $\calA = (V,V_P,E,T)$. 
Let $\calC$ be the target arena $(S,S_P,R,U)$ where
\begin{itemize}
    \item $S_P = \{ \eqclass{v} \st \exists v \in V_P \}$,
    \item $S = V_N \cup S_P$,
    \item $U = \{ \eqclass{t} \st \exists t \in T\}$, and
    \item $R = \{ (\eqclass{u},v)
        \st \exists u \in V_P \exists v \in V_N : (u,v) \in E \text{ and } vE
        \setminus \eqclass{u} \neq \emptyset\}
        \cup \{u,\eqclass{v}) \st \exists u \in V_N \exists v \in
        V_P : (u,v) \in E\}$.
\end{itemize}
For a family $\mu = (\mu_u \in \dist{uE})_{u \in V_N}$ of full-support
probability distributions we define $\lambda = (\lambda_u \in
\dist{uR})_{u \in V_N}$ as follows. For all $u \in V_N$ and all $\eqclass{v} \in
uR$ we have
\[ 
    \lambda_u(\eqclass{v}) = \sum_{w \in \eqclass{v}} \mu_u(w).
\]

\begin{proposition}\label{pro:prot-collapsing-safety}
    For all families $\mu = (\mu_u \in \dist{uE})_{u \in V_N}$ of full-support
    probability distributions and all $v \in V_P$ we have
    \[ 
        \max_\sigma \reachprob{\arenatochain{\calA}{\mu}{\sigma}}{v}{T} =
        \max_{\sigma'}
        \reachprob{\arenatochain{\calC}{\lambda}{\sigma'}}{\eqclass{v}}{U},
    \]
    where $U = \{\eqclass{t} \st t \in T\}$, and $\calC$ and $\lambda$ are as
    defined above.
\end{proposition}
\startmyproof
    For convenience, let us assume that
    extremal-probability vertices have already been collapsed so that for all 
    $u,v \in V_P$, if $u \sim v$ then $u$ and $v$
    are not extremal-probability vertices. The argument for correctness of
    collapsing extremal-probability vertices is trivial. We will reduce
    correctness of collapsing NWR-equivalence classes to the correctness of
    collapsing end components.
    
    Consider the following equation system defined for an arbitrary $\mu$.
    \begin{equation}\label{eqn:system1}
        \text{For all } u \in V_P, x_u = \begin{cases}
            1 & \text{if } u \in T\\
            0 & \text{if } u \in \zero{T}\\
            \max_{v \in uE} \sum_{w \in vE} \mu_v(w) \cdot x_w &
            \text{otherwise}
        \end{cases}
    \end{equation}
    According to Lemma~\ref{lem:bellman-ineqs}, the system has a unique solution
    assignment
    \begin{equation}\label{eqn:solution}
        x_u =
        \max_\sigma \reachprob{\arenatochain{\calA}{\mu}{\sigma}}{u}{T}
    \end{equation}
    for all $u \in V_P$.
    Now, by definition of the NWR we have that~\eqref{eqn:solution} is also a
    solution to the following system of equations. For all $u \in V_P$
    \begin{equation}\label{eqn:system2}
        \text{For all } u \in V_P, x_u = \begin{cases}
            1 & \text{if } u \in T\\
            0 & \text{if } u \in \zero{T}\\
            \max(\left\{ \sum_{w \in vE} \mu_v(w) \cdot x_w
                \:\middle|\: v \in uE \right\}\\
                \phantom{\max { }}
                \cup \left\{ x_y \:\middle|\: y \in \tilde{u} \right\}) &
                \text{otherwise}
        \end{cases}
    \end{equation}
    
    Observe that equation system~\eqref{eqn:system2} corresponds to the target
    arena $\calC$ we obtain by modifying $\calA$ as follows. For all $u \in V_P$
    and for all $v \in \tilde{u}$ we add an edge from $u$ to $v$ (with an
    intermediate Nature vertex to preserve bipartiteness). It thus follows from
    Lemma~\ref{lem:bellman-ineqs} that the system has a unique solution
    assignment corresponding to the maximal reachability probability values of
    all Protagonist vertices from $\calC$.
    We have already argued that~\eqref{eqn:solution} is a solution to the
    system~\eqref{eqn:system2}. Hence, since $\mu$ was chosen arbitrarily, we
    conclude that for all $u \in V_P$
    \[
        \max_\sigma \reachprob{\arenatochain{\calA}{\mu}{\sigma}}{u}{T} =
        \max_{\sigma'}
        \reachprob{\arenatochain{\calC}{\mu'}{\sigma'}}{u}{T},
    \]
    where $\mu'$ is the same as $\mu$ for all Nature vertices from $\calA$ and
    for all newly added nature vertices $w$, $\mu'_w$ assigns $1$ to the unique
    successor of $w$ in $\calC$.

    We now remark that in $\calC$ we have that for all $u \in V_P$ the set
    $\tilde{u}$ now forms an end component.
    Since collapsing end components
    preserves the desired value~\cite{dealfaro97,cbgk08} (and
    removing probability-$1$ self-loops clearly preserves this value too)
    the result follows.
\stopmyproof

We will now argue that collapsing NWR-equivalent Nature-owned vertices is also
safe. The following result will be useful.
\begin{lemma}\label{lem:nature-sim-successors}
    Consider the target arena $\calA$.
    For all $u,v \in V_N$ it holds that if $u \sim v$ and $u \neq v$ then $w
    \sim x$ for all $w,x \in (uE \cup vE)$.
\end{lemma}
\startmyproof
    We proceed by contradiction: suppose $u \sim v$, $u \neq v$, and that
    there exist $w,x \in (uE \cup vE)$ such that $w \not\sim x$. It follows from
    the latter that there are $w \in uE$ and $x \in vE$ such that $w \not\sim
    x$. Indeed, if this were not the case, then we would be able to conclude
    that $w \sim x$ for all $w,x \in (uE \cup vE)$ by using the transitivity of
    $\sim$. Without loss of generality, assume $\lnot(w \unlhd x)$. 
    
    We will now make use of the equation system~\eqref{eqn:system1} from the
    proof of Proposition~\ref{pro:prot-collapsing-safety} with $\mu$ the family
    that witnesses the fact that $\lnot(w \unlhd x)$, \ie $\val^\mu(w) >
    \val^\mu(x)$.  Recall that we have assumed $u \sim v$ and observe that
    \[
        x_w = \max_{z \in wE} \val^\mu(z),
    \]
    for all $w \not\in (T \cup \zero{T})$ in the
    system~\eqref{eqn:system1}. Now let $\mu'$ be a family of full-support
    probability distributions with $\mu'_u(w) = 1 - \epsilon$, where $0 <
    \epsilon < \frac{1}{2}(\val^\mu(w) - \val^\mu(x))$, and otherwise identical
    to $\mu$. Since $u \sim v$ we have that $\val^{\mu'}(u) = \val^{\mu'}(v) =
    \val^{\mu}(v) = \val^{\mu}(u)$ and thus by, Lemma~\ref{lem:bellman-ineqs},
    the maximal reachability probability values for all states in $\calA$ are
    the same for both $\mu$ and $\mu'$.  Similarly, we can obtain a family
    $\mu''$ such that $\mu''_v(x) = 1 - \epsilon$ without changing the values of
    the states in $\calA$. We thus have that
    \begin{align*}
        \val^{\mu''}(v) &\le \val^{\mu''}(x) + \epsilon & \\
        & < \val^{\mu''}(w) - \epsilon & \\
        & \text{since }
        \val^{\mu''}(x) = \val^{\mu}(x) < \val^{\mu}(w) =
        \val^{\mu''}(w) \text{ and by choice of } \epsilon & \\
        & \le (1 - \epsilon) \val^{\mu''}(w) & \\
        & \le \val^{\mu''}(u) &
    \end{align*}
    contradicting our initial assumption that $u \sim v$.
\stopmyproof

Let us now consider the following equation system --- that corresponds to
$\calC$ as defined above --- for an arbitrary $\mu$ and corresponding $\lambda$
as defined in the proof of Proposition~\ref{pro:prot-collapsing-safety}.
\begin{equation}\label{eqn:system3}
    \text{For all } \tilde{u} \in V_P, x_{\tilde{u}} = \begin{cases}
        1 & \text{if } \tilde{u} \in U\\
        0 & \text{if } \tilde{u} \in \zero{U}\\
        \max_{v \in \tilde{u}R} \sum_{\tilde{w} \in vR} \lambda_v(\tilde{w})
        \cdot x_{\tilde{w}} &
        \text{otherwise}
    \end{cases}
\end{equation}
It follows from Lemma~\ref{lem:nature-sim-successors} that for all $v \in V_N$
such that $|\tilde{v}| > 1$ the set $vR$ is a singleton $\{\tilde{w}\}$
and
\(
    \lambda_z(\tilde{w}) = 1.
\)
It is then easy to see that system~\eqref{eqn:system3} is the same equation
system that corresponds to $\calB$ as constructed for
Theorem~\ref{thm:collapsing}. The desired result thus follows from
Proposition~\ref{pro:prot-collapsing-safety} and Lemma~\ref{lem:bellman-ineqs}.
\qed

\section{Proof of Theorem~\ref{thm:opt-trim}}

The argument to show correctness for the removal of an edge to a sub-optimal
vertex is very similar to the proof of Theorem~\ref{thm:collapsing}. We
first consider the equation system of $\calA$ with an arbitrary $\mu$. Then,
we remove the variable from the maximum corresponding to the edge we would
remove in the target arena. The only way in which the latter would result in
the system not being the one corresponding to the reduced target arena
$\calB$ would be if by removing the edge, we make some vertex become an element
of $\zero{T}$. That is, we have removed all its paths to $T$. We will argue that
this is not possible if $\sim$-quotienting and self-loop removal
have already been applied.

We will need to formalize the intuition that if $\calA =
\quo{\calA}$ there are no cycles with probability $1$ in $\calA$.
\begin{lemma}\label{lem:no-prob1-cycles}
    Consider a target arena $\calA = (V,V_P,E,T)$ such that $\quo{\calA} =
    \calA$. For all families $\mu = (\mu_u \in \dist{uE})_{u \in V_N}$ of
    full-support distributions, and all $(v,w) \in E \cap (V_P \times V_N)$
    we have
    \[
        \max_\sigma \sum_{x \in wE} \mu_w(x)
        \reachprob{\arenatochain{\calA}{\mu}{\sigma}}{x}{v} < 1.
    \]
\end{lemma}
\startmyproof
    Towards a contradiction, suppose that there is some $\mu$, an edge
    $(v,w)$, and a strategy $\sigma$ for which
    \[
        \sum_{x \in wE} \mu_w(x)
        \reachprob{\arenatochain{\calA}{\mu}{\sigma}}{x}{v} = 1.
    \]
    Observe that from $v$, Protagonist can play $w$ and reach $wE$ almost
    surely. Also, modifying $\sigma$, to play $w$ from $v$ does not affect
    the probability of reaching $v$ from any $x \in wE$. Let $\sigma'$
    denote the modified strategy. It should be clear
    that $\sigma'$ is such that for all $p,q \in \{v\} \cup wE$
    \[
        \reachprob{\arenatochain{\calA}{\mu}{\sigma'}}{p}{q} = 1.
    \]
    Let $S :=\{v\} \cup wE$. If $S = \{v\}$ then we must have a
    probability-$1$ self-loop on $v$, which contradicts the fact that
    $\quo{\calA} = \calA$. Thus, $|S| \ge 2$. It follows that $S$ is a
    non-trivial end component and by Lemma~\ref{lem:probs-in-ec}, also an
    NWR-equivalence class. This, again, contradicts our initial assumptions.
\stopmyproof

We are now ready to prove our claim.
\startmyproof[of Theorem~\ref{thm:opt-trim}]
    We will only argue that $w$ in $\calB$ is not an element of $\zero{T}$.

    Let $Y:=wE \setminus \{x\}$. Observe $wE$ is non-empty. In fact, it must
    include at least $x$ and some other vertex so that $x \unlhd Y$ holds.
    Thus, since since $\quo{\calA} = \calA$, we have that $w \not\in
    \zero{T}$.

    Towards a contradiction, let us suppose that $w$ in
    $\calB = (V,V_P,E \setminus (w,x), T)$ is an element of $\zero{T}$. It
    follows that for all $s \in S:= \{s \in V_P \st \exists y \in Y : (y,s)
    \in E\}$ in $\calA$ every $s$--$T$ path goes through $w$. Hence, by
    Lemma~\ref{lem:sBt}, for all $\mu$ and for all $\sigma$ we have
    \begin{equation}\label{eqn:through-w}
        \reachprob{\arenatochain{\calA}{\mu}{\sigma}}{s}{T} = 
        \untilprob{\arenatochain{\calA}{\mu}{\sigma}}{s}{(Q \setminus \{w\})}{w}
        \reachprob{\arenatochain{\calA}{\mu}{\sigma}}{w}{T}.
    \end{equation}
    Let us now fix arbitrary $\mu$ and $y \in Y$ such that
    \[
        \val^\mu(w) = \sum_{z \in yE} \mu_y(z) \val^\mu(z).
    \]
    (The existence of such a $\mu$ and $y$ is guaranteed by
    Lemma~\ref{lem:bellman-ineqs} and the fact that $x \unlhd Y$.)
    We remark that $\val^\mu(x) \le \val^\mu(y)$. By definition of
    $\val^\mu(\cdot)$ we have that
    \[
        \val^\mu(y) = \sum_{z \in yE} \mu_y(z) \val^\mu(z).
    \]
    Together with Equation~\eqref{eqn:through-w} we obtain
    \[
        \val^\mu(y) =
        \max_\sigma
        \left(\reachprob{\arenatochain{\calA}{\mu}{\sigma}}{w}{T}
        \cdot
        \sum_{z \in yE} \mu_y(z)
        \reachprob{\arenatochain{\calA}{\mu}{\sigma}}{z}{w}\right).
    \]
    Finally, using Lemma~\ref{lem:no-prob1-cycles}, we conclude that
    $\val^\mu(y) < \val^\mu(w)$, contradicting our assumptions.
\stopmyproof

\section{Proof of Lemma~\ref{lem:probs-in-ec}}
\startmyproof
    Let $\sigma$ be a strategy maximizing the value
    $\reachprob{\calM^\sigma}{q}{T}$. We can then construct a (finite-memory)
    strategy which, from $p$ ensures $\untilprob{\calM^{\sigma'}}{p}{S}{q} = 1$
    and from $q$ onwards behaves as $\sigma$. If $q \in T$ then we are done.
    Otherwise, Lemma~\ref{lem:sBt} implies that
    $\reachprob{\calM^{\sigma'}}{p}{T} = \reachprob{\calM^{\sigma}}{q}{T}$.
    Furthermore, we know that memoryless strategies suffice for
    reachability~\cite{puterman05,bk08}, so
    $\sigma'$ can be replaced by a memoryless strategy.
\stopmyproof

\section{Proof of Lemma~\ref{lem:ndp}}
Let us denote by $\lambda$ the empty word.

\begin{definition}[Word simplification]
    For an alphabet $\Sigma$, we define the \emph{word simplification function}
    $s : \Sigma^* \to \Sigma^*$ as follows.
    \begin{itemize}
        \item $s(\lambda) := \lambda$
        \item $\forall w \in \Sigma^*,\forall a \in \Sigma:$
        \begin{itemize}
            \item If $a \notin s(w)$ then $s(wa) := s(w)a$,
            \item If $a \in s(w)$ then $s(wa)$ is the (shortest) prefix of
                $s(w)$ that ends with $a$.
        \end{itemize}
    \end{itemize}
\end{definition}

\begin{lemma}\label{lem:ws}
    Let $w$ be a word. $s(w)$ is repetition-free; $w$ and $s(w)$ have the same
    starting and ending letters; a letter occurs in $s(w)$ only if it occurs in
    $w$; a two-letter word $xy$ occurs as a factor in $s(w)$ only if it occurs
    in $w$.
\end{lemma}
\startmyproof
    By induction. The base case is clear. For the inductive case, let $w \in
    \Sigma^*$ and let $a \in \Sigma$.

    First case, $a \notin s(w)$. So $s(wa) = s(w)a$ is repetition-free like
    $s(w)$ is by IH, and $wa$ and $s(wa)$ share the same ending letter $a$.
    Since $w$ and $s(w)$ share the same starting letter by IH, so do $wa$ and
    $s(wa) = s(w)a$. (It is $a$ if $w = \lambda$). Let $b \in s(wa)$. If $b =
    a$, the claim clearly holds; if $b \neq a$ then $b \in s(w)$, so $b \in w$
    by IH, and $b \in wa$. Let $xy$ occur as a two-letter factor in $s(wa)$. If
    $y \neq a$ then $xy$ actually occurs in $s(w)$, so by IH it occurs in $w$,
    and thus in $wa$; if $y = a$ then $s(w)$ ends with $x$, and by IH so do $w$,
    so $wa$ ends with $xy$. 

    Second case, $a \in s(w)$. Since $s(wa)$ is a prefix of $s(w)$, it is also
    repetition-free, and $s(wa)$ ends with $a$ by definition. As a prefix,
    $s(wa)$ starts with the same letter as $s(w)$, \ie~the same letter as $w$ by
    IH, \ie~the same as $wa$. If $b \in s(wa)$ then $b \in s(w)$, so $b \in w$
    by IH, and $b \in wa$. If $xy$ occurs in $s(wa)$, it also occurs in its
    extension $s(w)$, and in $w$ by IH, and finally in $wa$.
\stopmyproof

\begin{lemma}\label{lem:path}
    For all paths $\gamma$, $s(\gamma)$ is a simple path starting (ending) with
    the same vertex as $\gamma$. It only visits vertices that are already
    visited in $\gamma$, and it only takes edges that are already taken in
    $\gamma$. 
\end{lemma}
\startmyproof
    By Lemma~\ref{lem:ws} the word $s(\gamma)$ is repetition-free, and it starts
    and ends like $\gamma$. Again by Lemma~\ref{lem:ws}, if two vertices $xy$
    occur consecutively in $s(\gamma)$, they also occur consecutively and in the
    same order in $\gamma$, \ie~$(x,y) \in E$. Therefore $s(\gamma)$ is a
    path, and moreover it takes only edges that are already taken in $\gamma$. 
\stopmyproof

We can now proceed with the proof of the Lemma.
\startmyproof[of Lemma~\ref{lem:ndp}]
    Since there is a path from $x$
    to $T$ by assumption, the set $L$ of all the \emph{simple} paths
    from $x$ to $T$ is non-empty. Let $\mathcal{T}$ be the prefix
    closure of $L$, and let $\mathcal{T}'$ be the set of the
    $\mu$-non-decreasing paths in $\mathcal{T}$, so that $\mathcal{T}' \subseteq
    \mathcal{T}$. The set $\mathcal{T}$ is a tree by prefix-closure
    construction, and $\mathcal{T}'$ is a tree since the $\mu$-non-decreasing
    paths are closed by taking prefixes. Moreover, the elements of $\mathcal{T}$
    (and thus $\mathcal{T}'$) are simple paths since the prefixes of simple
    paths are again simple paths. Let $L'$ be the prefix-wise maximal paths in
    $\mathcal{T}'$. The claimed lemma amounts to $L \cap L' \neq \emptyset$,
    equivalently $L \cap \mathcal{T}' \neq \emptyset$. Also note that a path in
    $\mathcal{T}$ is in $L$ if and only if it ends with $T$, so $L
    \cap \mathcal{T}' \neq \emptyset$ if and only if some path in $\mathcal{T}'$
    ends with $T$.

    Towards a contradiction, let us assume that $L \cap L' = \emptyset$. We can
    thus let $M$ consist of the one-vertex extensions in $\mathcal{T}$ of the
    paths in $L'$. More formally, a path $\gamma\in \mathcal{T}$ is in $M$ if
    and only if there exists a vertex $y$ and a path $\beta \in L'$ such that
    $\beta y = \gamma$.  Clearly, $\mathcal{T}' \cap M = \emptyset$ by
    prefix-wise maximality of $L'$ within $\mathcal{T}'$. So, the elements in
    $M$ are the least \emph{not} $\mu$-non-decreasing paths in $\mathcal{T}$.
    This implies that 
    \begin{align}
        \forall \gamma yz \in M,\quad \val^\mu(z) < \val^\mu(y)
        \label{eqn:val1}
    \end{align}
    (where $\gamma$ is a path and $y$, $z$ are vertices), since $\gamma y \in L'$.
    Let $B'$ (resp. $B$) be the ending vertices of the paths in $L'$ (resp. $M$).
    Let us proceed with a few remarks:

    \begin{remark}\label{rem:proof}
        \begin{enumerate}
            \item\label{rem:proof1} $x$ occurs as the first vertex of all paths
                in $\mathcal{T}$, and only as first vertex, otherwise the paths
                would not be simple. So $x \notin B$, since $x$ (as a path) is
                clearly in $\mathcal{T}'$ and since $\mathcal{T}' \cap M =
                \emptyset$.

            \item\label{rem:proof2} Every path in $L$ has a prefix in $L'$
                (\textit{i.e.} its longest $\mu$-non-decreasing prefix), and
                it has a prefix in $M$ too, since $L \cap L' = \emptyset$. So
                every path in $L$ visits $B$.

            \item\label{rem:proof3} Every path from $x$ to $T$ visits
                $B$: indeed, if a path $\gamma$ from $x$ to $T$ does
                not visit $B$, neither does $s(\gamma)$ by Lemma~\ref{lem:path}.
                Again by Lemma~\ref{lem:path}, $s(\gamma)$ is a simple path
                starting at $x$ and ending at $T$, so $s(\gamma) \in
                L$, so it must visit $B$ by
                Remark~\ref{rem:proof}.\ref{rem:proof2}, contradiction  
        \end{enumerate}
    \end{remark}

    Let us now prove that for all paths $\gamma \in \mathcal{T}'$ starting with
    $x$, and for all paths $\beta$, if $\gamma \beta$ is a path ending with
    $T$, then $\beta$ visits $B$ (even if $\gamma \beta \notin
    \mathcal{T}$). We proceed by induction on $\gamma$. Base case, $\gamma = x$
    is a one-vertex path: since $x \notin B$ by
    Remark~\ref{rem:proof}.\ref{rem:proof1}, $\beta$
    visits $B$ for all paths $x\beta$ ending with $T$, by
    Remark~\ref{rem:proof}.\ref{rem:proof3}.

    For the inductive case, let $\gamma'y \in \mathcal{T}'$ and let us assume
    that the claim holds for the prefixes of $\gamma'$. Let $y\beta'$ be a path
    where
    $\beta'$ ends with $T$. By Lemma~\ref{lem:path}, $s(y\beta')$ can
    be decomposed as $y\beta$ for some $\beta$. By Lemma~\ref{lem:path} again,
    every vertex occurring in $\beta$ also occurs in $\beta'$, and $y\beta$ is a
    simple path ending in $T$. Let us make a case disjunction. First
    case, $\gamma'y\beta$ is simple, so it is in $L$, and it has a prefix in $M$
    by Remark~\ref{rem:proof}.\ref{rem:proof2}. Since $\mathcal{T}' \cap M =
    \emptyset$, this prefix is a proper extension of $\gamma'y \in
    \mathcal{T}'$, so $\beta$ visits $B$. Second case, $\gamma'y\beta$ is not
    simple. Since $\gamma'y$ and $y\beta$ are simple, some vertex $z$ in
    $y\beta$ occurs already in $\gamma'$, hence the following decompositions:
    $\gamma' = \gamma_\ell z\gamma_r$ and $y\beta = \beta_\ell z\beta_r$. The
    path $\gamma_\ell z$ is in $\mathcal{T}'$, the path $\gamma_\ell z\beta_r$
    ends with $T$, so by induction hypothesis $\beta_r$ visits $B$,
    and so does $\beta$, and so does $\beta'$
    again by Lemma~\ref{lem:path}. The induction is thus complete.

    On the one hand, let $\gamma y \in L'$ be such that $\val^\mu(y)=
    \max\{\val^\mu(z) \st z \in B'\}$. So $\val^\mu(z) <
    \val^\mu(y)$ for all $z \in B$ by (\ref{eqn:val1}).

    On the other hand, let $\sigma_0$ be an arbitrary positional strategy for
    Protagonist in the MDP. Every path from $y$ to $T$ visits $B$, as
    proved by induction above, which also holds in the graph restricted by
    $\sigma_0$, so $\untilprob{\arenatochain{\calA}{\mu}{\sigma_0}}{y}{\left(V
    \setminus B\right)}{T} = 0$.
    Then Lemma~\ref{lem:sBt} implies 
    \begin{align*}
        \reachprob{\arenatochain{\calA}{\mu}{\sigma_0}}{y}{T}
        & =
        \sum_{z \in B}
        \untilprob{\arenatochain{\calA}{\mu}{\sigma_0}}{y}{\left(V \setminus
        B\right)}{z}\reachprob{\arenatochain{\calA}{\mu}{\sigma_0}}{z}{T}\\
        & \leq
        \sum_{z \in B}
        \untilprob{\arenatochain{\calA}{\mu}{\sigma_0}}{y}{\left(V \setminus
        B\right)}{z}
        \max_\sigma
        \reachprob{\arenatochain{\calA}{\mu}{\sigma_0}}{z}{T}\\
        & =
        \sum_{z \in B}
        \untilprob{\arenatochain{\calA}{\mu}{\sigma_0}}{y}{\left(V \setminus
        B\right)}{z}
        \val^{\mu}(z) \\
        & \leq
        \sum_{z \in B}
        \untilprob{\arenatochain{\calA}{\mu}{\sigma_0}}{y}{\left(V \setminus
        B\right)}{z}
        \max_{b \in B}
        \val^{\mu}(b) \\
        &  \leq \max_{b \in B}\val^{\mu}(b), \text{ since }
        \sum_{z \in B}
        \untilprob{\arenatochain{\calA}{\mu}{\sigma_0}}{y}{\left(V \setminus
        B\right)}{z} \leq 1.
    \end{align*}
    Therefore, $\val^\mu(y) =
    \max_{\sigma}\reachprob{\arenatochain{\calA}{\mu}{\sigma}}{y}{T}
    \leq  \max_{b \in B}\val^{\mu}(b)$.  This
    contradicts the above claim that $\val^\mu(z) < \val^\mu(y)$ for all
    $z \in B$.
\stopmyproof

\section{Proof of Proposition~\ref{pro:bar-rules}}
The following observation regarding convex combinations will be useful.
\begin{lemma}\label{lem:convex-combo}
    Consider a finite set of values $N \subseteq \mathbb{Q}$ and a probability
    distribution $\delta \in \dist{N}$. There exist $n,m \in N$ such that
    \begin{itemize}
        \item $\exists \underline{m} \in N: \underline{m} \leq \sum_{n \in N}
            \delta(n) \cdot n$
        \item $\exists \overline{m} \in N: \sum_{n \in N} \delta(n) \cdot n \leq
            \overline{m}$
    \end{itemize}
\end{lemma}

\startmyproof[of Proposition~\ref{pro:bar-rules}]
    Let us start with item~\ref{itm:happy-bar}. From Lemma~\ref{lem:sBt} we have
    that
    \[
        \reachprob{\arenatochain{\calA}{\mu}{\sigma}}{v_0}{T} = \sum_{s \in S}
        \untilprob{\arenatochain{\calA}{\mu}{\sigma}}{v_0}{\left(V_P \setminus
        S\right)}{s}
        \reachprob{\arenatochain{\calA}{\mu}{\sigma}}{s}{T}
    \]
    for all families $\mu$ of full-support distributions and strategies
    $\sigma$. It follows from Lemma~\ref{lem:convex-combo} that there is some $s
    \in S$ such that 
    \[
        \reachprob{\arenatochain{\calA}{\mu}{\sigma}}{v_0}{T}\le
        \reachprob{\arenatochain{\calA}{\mu}{\sigma}}{s}{T}.
    \]
    By definition, this
    means that $v_0 \unlhd S$. Hence, by choice of
    $S$ and Remark~\ref{rem:props-trans-closure} we have $v_0 \unlhd W$.

    For item~\ref{itm:sad-bar}, let us assume, without loss of generality, that
    the vertices from $T\cup \zero{T}$ are all sinks.
    We thus have that there is a strategy $\sigma$ such
    that in there is no path from $v_0$ to $\zero{T}$
    without first visiting $T \cup S$.
    We will also assume that $\sigma$, after visiting any
    vertex from $S \cup T$ starts playing optimally in order to maximize the
    probability of visiting $T$. (This may require memory.)
    We have from Lemma~\ref{lem:sBt} that
    \[
        \reachprob{\arenatochain{\calA}{\mu}{\sigma}}{v_0}{\zero{T}} =
        \sum_{s \in S
        \cup T}
        \untilprob{\arenatochain{\calA}{\mu}{\sigma}}{v_0}{V_P \setminus (T \cup
        S)}{s}
        \reachprob{\arenatochain{\calA}{\mu}{\sigma}}{s}{\zero{T}}
    \]
    for all full-support distribution families $\mu$. Since all the vertices
    from $T$ are
    sinks
    $\reachprob{\arenatochain{\calA}{\mu}{\sigma}}{s}{\zero{T}} = 0$ for all $s
    \in T$. Therefore, we can rewrite the above as
    \[
        \reachprob{\arenatochain{\calA}{\mu}{\sigma}}{v_0}{\zero{T}} =
        \sum_{s \in S}
        \untilprob{\arenatochain{\calA}{\mu}{\sigma}}{v_0}{V_P \setminus (T \cup
        S)}{s}
        \reachprob{\arenatochain{\calA}{\mu}{\sigma}}{s}{\zero{T}}.
    \]
    From
    Lemma~\ref{lem:convex-combo} we get that there is some $s \in S$ such
    that 
    \begin{align*}
        &\reachprob{\arenatochain{\calA}{\mu}{\sigma}}{v_0}{\zero{T}} \leq
        \reachprob{\arenatochain{\calA}{\mu}{\sigma}}{s}{\zero{T}}&\\
        \iff &
        1 - \reachprob{\arenatochain{\calA}{\mu}{\sigma}}{v_0}{T} \leq
        1 - \reachprob{\arenatochain{\calA}{\mu}{\sigma}}{s}{T}
        & \sigma \text{ is maximizing the prob. of reaching } T
        \\
        \iff &
        \reachprob{\arenatochain{\calA}{\mu}{\sigma}}{s}{T} \leq
        \reachprob{\arenatochain{\calA}{\mu}{\sigma}}{v_0}{{T}}.&
    \end{align*}
    By choice of $\sigma$, this implies that $s \unlhd \{v_0\}$ (recall that
    memoryless strategies suffice to maximize the probability of reaching a
    target set of states). Then, by choice of $S \ni s$ we have that $w \unlhd
    \{v_0\}$.
\stopmyproof

\section{Proof of Proposition~\ref{pro:equiv-rules}}

\startmyproof
    Item~\ref{itm:equiv-nature} follows from the definition of $\val$.

    For item~\ref{itm:equiv-prot}, we suppose the assumptions hold. Hence, by
    definition of $\unlhd$, we have that for all families $\mu$
    of full-support distributions
    \[
        \max_{w \in uE} \sum_{x \in wE} \mu_w(x) \val^\mu(x) \le \max_{y \in vE}
        \sum_{z \in yE} \mu_y(z) \val^\mu(z).
    \]
    The result thus follows from Lemma~\ref{lem:bellman-ineqs}.
\stopmyproof

\end{document}